# Cold plasma with zirconia nanoparticles for lung cancer via TGF-β signaling pathway

Yueye Huang[1,2,#], Rui Zhang[3,#], Xiao Chen[1], Fei Cao[1], Qiujie Fang[1,2,4], Qingnan Xu[1], Shicong Huang[2], Yufan Wang[5], Guojun Chen[4,6,*], Zhitong Chen[1,2,7,*]

[1]Paul C Lauterbur Research Center for Biomedical Imaging, Institute of Biomedical and Health Engineering, Shenzhen Institute of Advanced Technology, Chinese Academy of Sciences, Shenzhen, 518055, China

[2]Advanced Therapeutic Center, National Innovation Center for Advanced Medical Devices, Shenzhen 518000, China

[3]Department of Prosthodontics, Stomatology Center, Peking University Shenzhen Hospital, Shenzhen 518036, China

[4]Department of Biomedical Engineering, McGill University, Montreal, QC, H3G 0B1, Canada

[5]Department of Oral and Maxillofacial Surgery, Peking University Shenzhen Hospital, Shenzhen 518036, China

[6]Rosalind & Morris Goodman Cancer Institute, McGill University, Montreal, QC, H3G 0B1, Canada

[7]Key Laboratory of Biomedical Imaging Science and System, Chinese Academy of Sciences, Shenzhen, 518055, China

*Corresponding author. Email: zt.chen1@siat.ac.cn (Z.C.), guojun.chen@mcgill.ca (G.C.); #Y.H. and R.Z. contributed equally to this work.

## ABSTRACT

Despite advancements in lung cancer therapy, the prognosis for advanced or metastatic patients remains poor, yet many patients eventually develop resistance to standard treatments leading to disease progression and poor survival. Here, we described a combination of CAP and nanoparticles ($ZrO_2$ NPs (zirconium oxide nanoparticle) and 3Y-TZP NPs (3% mol Yttria Tetragonal Zirconia Polycrystal Nanoparticle)) for lung cancer therapy. We found that $ZrO_2$ NPs caused obvious damage to the inside of the lung cancer cells. CAP and $ZrO_2$ NPs mainly affected the mitochondria function, leading to a decrease in mitochondrial membrane potential and ATP levels, and causing endoplasmic reticulum stress and cell nucleus internal DNA damage, etc. CAP combined with $ZrO_2$ NPs (CAP@$ZrO_2$) induced lung cancer cell apoptosis by activating the TGF-β pathway. CAP@$ZrO_2$ offers a new therapy for the clinical treatment of lung cancer.

## INTRODUCTION

Non-small cell lung cancer (NSCLC) is the most prevalent type of lung cancer, accounting for around 90% of total cases.[1, 2] It is typically not diagnosed until the late stages due to the mild nature and non-specificity of early symptoms. This late diagnosis often results in over 50% of NSCLC patients dying within a year after diagnosis and only 20.5% surviving 5 years post-diagnosis.[3, 4] As current treatments are limited, novel therapeutic approaches are needed. Cold atmospheric plasma (CAP), operating at atmospheric pressure and room temperature, can produce selective sensitization or selective cell death without affecting normal cells.[5-9] CAP delivers reactive oxygen species (ROS) and reactive nitrogen species (RNS) as well as multiphysical fields.[10-15] CAP effects on cancer cells are caused by either ROS/RNS or multiphysical fields.[16-19] Compared with conventional therapies, such as radiotherapy and chemotherapy, CAP technology employs ionized gas for cancer treatment without additional substances.[13, 20-22] Plasma induces selective cancer cell death due to cancer cells having a higher basal oxidative level than normal cells due to hypermetabolism.[23-25] The phase I clinical trial of CAP in combination with surgical tumor resection has demonstrated the improvement of the overall survival rate for patients even with stage IV solid tumors.[26]

In the past few years, nanoparticles have acquired the strong attention of the scientific community and have been thoroughly investigated in the field of human health for tumor imaging, cytotoxic activity, and drug delivery.[27-29] Several metal oxides such as zinc oxide, copper oxide, zirconium oxide, manganese oxide, iron oxide, and others play a significant role in various medical applications.[30-32] The low toxicity and high chemical inertness of zirconium oxide nanoparticles ($ZrO_2$ NPs) consider them environmentally friendly materials as well as prospective biological agents.[33, 34] $ZrO_2$ NPs possess several favorable characteristics for anticancer drug delivery with improved therapeutic efficacy and the potential application of imaging for real-time monitoring of the diagnosis and treatment process.[35] $ZrO_2$ NPs cause cytotoxicity in a number of cancer cells due to the ROS pathway.[36] Appropriate concentration of $ZrO_2$ NPs induced Intracellular ROS of cancer cells.[37] Kadiyala et al. reported $ZrO_2$ NPs with distinct anticancer effects toward human HCT116 and A549 cancer cell lines. Their cytotoxic effect is more pronounced and mediated via oxidative stress by generating intracellular ROS in a manner similar to that of CAP, which can induce/trigger apoptosis.[38] Whether combine CAP and $ZrO_2$ NPs to enhance the killing effects provides great potential for lung cancer therapy.

In this paper, we report a novel and promising therapy via combining CAP and nanoparticles ($ZrO_2$ NPs and 3Y-TZP NPs) for lung cancer (Fig. 1). Firstly, we studied its cytotoxic effects by MTT assay and Calcein-AM/PI double labeling method. In addition, we investigated ROS and RNS levels in the culture medium under different treatments, and demonstrated the effect on mitochondrial membrane potential and intracellular ATP content. Furthermore, we conducted transcriptome analysis and obtained differential genes that were significantly up-regulated and down-regulated under different experimental conditions. Through different enrichment analysis methods, we learned about the signaling pathways and biological processes involved in different treatment conditions, which helped reveal the molecular regulatory network of lung cancer cells and the impact of specific treatments on cell function and metabolism. Our results are of great significance for further studying the mechanisms of lung cancer development, metastasis, and drug resistance. This novel therapeutic approach (CAP@$ZrO_2$) holds great potential for lung cancer clinical therapy.

## EXPERIMENTAL SECTION

*Cells*

A549, a human lung adenocarcinoma cell line was cultured in F-12K medium with L-glutamine (Meisen CTCC, Zhejiang, China) supplemented with 10% fetal bovine serum and 1% Penicillin-Streptomycin Solution (Procell, WuHan, China) in a humidified atmosphere of 5% $CO_2$ at 37 ℃. ZrO2 and 3Y-TZP powders are purchased from Sinocera (Jiangsu, China).

*Plasma device and diagnostics*

The plasma jet device employed in this study is composed of the metal electrode rod (2 mm diameter) and the ground electrode (a ring-shaped copper foil) ( Fig.S1). Thermal images of the CAP device were taken using a handheld thermography camera (HIKMICRO, HM-TPH21Pro-3AQF). The discharge voltage was measured by an oscilloscope (Tektronix, TDS2024C) with a high-voltage probe (Tronovo, TR9340A). The optical emission spectrometry (OES) was characterized by a high-resolution UV/visible spectrometer (Brolight, BIM-6602A series), setting up the optical probe placed 1 cm away from the discharge area. The discharge process of CAP was recorded by a time-resolved image enhancement camera (Intelligent Scientific System, TRC411).

*TEM and XRD*

The Scanning Electron Microscope (SEM) analysis was conducted using an Ultra-high resolution field emission scanning electron microscope (Zeiss, Gemini300). To ensure accurate electronic signals and high reflectivity for non-conductive materials, $ZrO_2$ and 3Y-TZP surfaces were coated with a conductive layer of gold. Subsequently, the topographical features of CAP-treated and non-treated $ZrO_2$ and 3Y-TZP surfaces were observed at high resolution, encompassing surface texture, shape, roughness, and pore structure. For deeper insights into the morphological characteristics of $ZrO_2$ and 3Y-TZP, a Talos F200X Field Emission Transmission Electron Microscope from FEI, USA, was utilized. This enabled nanoscale microanalysis and facilitated the analysis of the crystalline structure and lattice parameters of $ZrO_2$ and 3Y-TZP. High-resolution lattice streak images were employed to scrutinize the crystal structures and lattice parameters further. Moreover, an X-ray diffractometer (Bruker, D2 PHASER) was employed to investigate the phase composition of the nanoparticles. This examination allowed for the observation of grain structure at the microscopic level, with the phase composition of the substance identified by comparing the diffraction pattern obtained with a standard PDF card. The software was utilized to eliminate interfering data, aiding in the accurate determination of the phase composition.

*Fourier infrared spectroscopy detection*

The specific quantity of $ZrO_2$/ 3Y-TZP was dispersed in deionized water and thoroughly mixed using a mixer (Thermo, 88882010). Subsequently, 200 μL of the resulting mixture was drawn up with a pipette gun and applied onto a disinfected slide, then placed in a dryer (Shanghai Yihang, DHG-9075A) to eliminate excess deionized water. Upon drying, the zirconia nanoparticles were retrieved, and their surfaces were subjected to plasma treatment for durations of 10 s, 20 s, 30 s, 60 s, and 120 s, respectively, completing the pre-sample preparation. A Fourier Infrared Spectrometer (Thermo Fisher, Nicolet iS20) was employed to examine the impact of plasma treatment on the surface chemical composition and functional groups of $ZrO_2$/ 3Y-TZP over different time intervals. For sample preparation, 1-2 mg of zirconia nanoparticles were ground into a fine powder using an onyx mortar, then mixed with dry potassium bromide (A.R. grade) powder (approximately 100 mg, 200 mesh particle size). The mixture was loaded into molds and compressed with a pressure not exceeding 20 mPa using a tablet press. The resulting tablets were thoroughly mixed, loaded into molds, and pressed into tablets on a tablet press with a pressure not exceeding 20 mPa for subsequent testing. In this study, powdered nanoparticles were the subject of examination, with the band range selected from 4000 to 400 $cm^{-1}$, precision of 4 $cm^{-1}$, and 32 scans conducted for each measurement, repeated in triplicate for accuracy. The zeta potential of two types of zirconia nanoparticles was measured using a particle size meter (Bruker, Nanobrook Omni).

*Hydrated particle sizes*

The hydrated particle sizes of two types of $ZrO_2$/3Y-TZP were determined using a particle size meter (Bruker, Nanobrook Omni). A specific quantity of $ZrO_2$/3Y-TZP was dispersed in F-12K medium solution, respectively. Utilizing the principle of dynamic light scattering, the equivalent particle size of the nanoparticles was measured. For the experimental procedure, 2 mL of the prepared samples were transferred into a cuvette. The instrument parameters were set to a 90-degree angle, with a single measurement time of 3 minutes and a measurement interval of 1 minute. Each measurement was conducted in triplicate to ensure accuracy. The particle size distribution and polydispersity of the zirconia nanoparticles were determined by analyzing the measured values of dynamic light scattering using the BIC Particle Solutions software (version 3.6.0) correlation function.

*Intracellular esterase activity and cell membrane integrity*

Calcein-AM (calcein-acetoxymethyl) and Propidium Iodide (PI) are utilized to assess intracellular esterase activity and cell membrane integrity, respectively, to detect cell viability and cytotoxicity. The combination of Calcein AM and PI enables simultaneous double fluorescence staining of live and dead cells. In this assay, the effects of various treatments, including CAP (2 minutes), $ZrO_2$ (300 μg/mL), 3Y-TZP (300 μg/mL), or their combination, on cell viability and toxicity were evaluated. An automated inverted fluorescence microscope (Nikon, ECLIPSE Ti2-E) was employed to observe and analyze the impact on cell activity and toxicity.

*Cell viability*

Cell viability was assessed using the MTT assay (Solarbio, M1020). The mitochondrial membrane potential was assessed using a cation probe TMRE. To evaluate apoptosis or necrosis based on changes in fluorescence signal intensity, cells were examined using an automated inverted fluorescence microscope (Nikon, ECLIPSE Ti2-E) and a multimodal microplate reader (TECAN, Infinite 200 Pro), which enable the quantification and analysis of fluorescence signals to assess mitochondrial membrane potential and cellular health status. The ATP concentration assay was conducted using a multimodal microplate reader (TECAN, Infinite 200 Pro) capable of detecting chemiluminescence.

*ROS and RNS*

A hydrogen peroxide assay kit (Beyotime, S0038) was utilized to quantify the concentration of $H_2O_2$ in various mediums treated with CAP, $ZrO_2$, 3Y-TZP, and their combination, following the manufacturer's instructions. The absorbance of each well was measured at 560 nm using a multimodal microplate reader (TECAN, Infinite 200 Pro). A NO detection kit (Beyotime, S0021M) was used to measure the level of nitrates (NOx-) generated in mediums treated with CAP, $ZrO_2$, 3Y-TZP, and their combination, following the manufacturer's instructions. After the reaction period, the absorbance of each well was measured at 540 nm using a multimodal microplate reader (TECAN, Infinite 200 Pro). The intracellular ROS and RNS levels were measured using the Reactive Oxygen Species Assay Kit (S0033M, Beyotime Biotechnology, China) and the Reactive Nitrogen Species Assay Kit (S0019M, Beyotime Biotechnology, China), respectively. Changes in fluorescence intensity before and after stimulation were detected using a flow cytometer with an excitation wavelength of 488 nm and an emission wavelength of 525 nm for ROS detection. For RNS detection, the fluorescent dye DAF-FM diacetate at 5 μM was used, and fluorescence was detected at an excitation wavelength of 495 nm and an emission wavelength of 515 nm.

*RT-qPCR*

Total RNA was extracted by Trizol (Vazyme, Nanjing, China), phase-separated with chloroform, precipitated with isopropyl alcohol, washed with 75% ethanol, and re-dissolved in water. Reverse transcription was performed with 1 μg RNA and random hexamers using the PrimeScript™ RT reagent Kit (RR047A; Takara, Beijing), China). RT-qPCR primers (Table 1) were designed by Primer3. RT-qPCR was conducted to validate transcriptome sequencing outcomes on the LightCycler® 96 Instrument(Roche, CA, Switzerland) utilizing SYBR reagents(807302X, Dakewe, Shenzhen, China).

Table 1 RT-qPCR primers

| Gene | Forward | Reverse |
|---|---|---|
| TGF-β | GACTTTTCCCCAGACCTCGG | ATAGGGGATCTGTGGCAGGT |
| Smad2 | GCTGGCCTGATCTTCACAGT | CCAGAGGCGGAAGTTCTGTT |
| Smad3 | GTTGGTGGAGGGTGTAGTGG | GGTTCAGAGGACCCTTGTGG |
| Smad4 | GCTGCAGAGCCCAGTTTAGA | GGCTCTTCTCTGGCTTCTGG |
| Bcl-2 | GAACTGGGGGAGGATTGTGG | CATCCCAGCCTCCGTTATCC |
| Bax | AAACACAGTCCAAGGCAGCT | AAACACAGTCCAAGGCAGCT |
| GAPDH | GTCAAGGCTGAGAACGGGAA | AAATGAGCCCCAGCCTTCTC |

A549 cells in the logarithmic growth phase were plated in six-well plates and treated with CAP (2 minutes), $ZrO_2$ (300 μg/mL), 3Y-TZP (300 μg/mL), and their combination for 24 hours. RNA samples were extracted using Trizol reagent (Invitrogen, CA, USA) following the manufacturer's protocol. The purity and quantification of RNA were evaluated using the NanoDrop 2000 spectrophotometer (Thermo Scientific, USA), while RNA integrity was assessed using the Agilent 2100 Bioanalyzer (Agilent

Technologies, Santa Clara, CA, USA). Subsequently, RNA libraries were constructed using the VAHTS Universal V6 RNA-seq Library Prep Kit following the manufacturer's instructions. The libraries were then subjected to transcriptome sequencing and analysis by OE Biotech Co., Ltd. (Shanghai, China) using high-throughput RNA sequencing technologies. The bioinformatic analysis of the RNA sequencing data involved various steps, including quality control, read alignment to a reference genome or transcriptome, quantification of gene expression levels, identification of differentially expressed genes, functional enrichment analysis, and pathway analysis.

*HTSeq-count*

The Fragments Per Kilobase of transcript per Million mapped reads (FPKM) of each gene were calculated, and the read counts of each gene were obtained using HTSeq-count. Principal Component Analysis (PCA) was conducted using R (v 3.2.0) to assess the biological duplication of samples. Differential expression analysis was performed using DESeq2, with a threshold of Q value $< 0.05$ and fold change $> 2$ or fold change $< 0.5$ set for significantly differentially expressed genes (DEGs). Hierarchical cluster analysis of DEGs was performed using R (v 3.2.0) to illustrate the expression patterns of genes across different groups and samples. Additionally, a radar map of the top 30 genes was generated to visualize the expression levels of up-regulated or down-regulated DEGs using the R package grader.

*Statistical analysis*

One-way ANOVA with Tukey post-hoc tests (multiple comparisons) was performed to determine statistical significance between different groups. Data points represent mean ± standard deviation. All experiments were repeated independently with similar results at least three times. All statistical analyses were performed using the software GraphPad Prism version 8 (GraphPad Software Inc.). Significance was denoted in Figures as ns, nonsignificant ($P > 0.05$), *$P < 0.05$, **$P < 0.01$, ***$P < 0.001$, and ****$P < 0.0001$.

## RESEAULTS AND DISCUSSION

Fig. 2a and Fig. S2 depict the discharge evolution process of the plasma within one discharge cycle. It can be observed that at the beginning of plasma discharge, the plasma moves irregularly, forming discontinuous bright spots known as plasma bullets. These bright spots move rapidly along the length of the jet during the spraying process. Plasma jets produce at least one prominent plasma bullet within each voltage cycle. The formation, collapse, and regeneration of these bullets occur repeatedly near the cathode. The presence of these bullets indicates the periodic nature of the plasma discharge process, with similar behavior repeated within each cycle.

The temperature of the plasma jet device ranges from 17.8 to 32.1°C (Fig 2b), indicating that the plasma temperature remains at room temperature. The input voltage and current of the device are 7.50V and 0.60A, respectively, while the helium gas flow rate is 7.5 L/min. Fig. 2c shows the waveform of the plasma discharge voltage. When the power supply is turned on without connecting the plasma jet device, the peak-to-peak output voltage and frequency are approximately 9.44 kV and 17.52 kHz, respectively. After connecting the helium jet plasma device, the peak-to-peak output voltage decreases to approximately 8.16 kV. When a 96-well plate containing culture medium is placed beneath the jet, the discharge voltage further decreases to 7.48 kV, the reason for this decrease in voltage could be attributed to the interaction of the plasma jet with the liquid surface of the culture medium. This interaction may induce ionization reactions, charge transfer, and interfacial chemical reactions in the liquid, leading to an increase in the degree of ionization and conductivity of the liquid.[39, 40] As a result, the voltage drops across the liquid when current passes through it decreases. Fig. 2d displays the relative intensity of OES ranging from 200-1000nm, detecting the emission spectrum of the plasma jet with and without contact with the culture medium. The peaks in the spectrum indicate the presence of RNS and ROS in the helium plasma jet generated by this device. Specifically, peaks at 391.50nm and 428.51nm correspond to the photon emission intensity during $N_2$ excitation, representing $N_2$ and $N_2^+$ respectively. Additionally, the wavelength at 777.64nm can be defined as O.[41]

Fig. 3a shows that the diffraction peak signals of $ZrO_2$ NPs are prominent, corresponding to the (110), (002), (111), and (300) crystal planes. These diffraction peaks are characteristic peaks of monoclinic-phase zirconia in X-ray diffraction (XRD), with the monoclinic phase content being approximately 95%. The main crystal diffraction peaks of 3Y-TZP NPs samples are observed at (011), (110), (112), and (211), which are characteristic peaks of tetragonal-phase zirconia in XRD. The results did not reveal the presence of a second phase, indicating no impurity peaks. This suggests that the yttrium elements in the raw material are completely solid-solubilized into the lattice of zirconia, thereby stabilizing the zirconia.

The high-resolution electron microscope image in Fig. 3b can distinguish that the (110) crystal plane and the (111) crystal plane belong to a single Oblique phase $ZrO_2$, and (101) crystal plane belongs to the tetragonal phase $ZrO_2$. Most of the grains in Fig. 3b and Fig.S3 have parallel and equidistant lattice stripes, and there is no lattice mismatch, indicating that the sample is well crystallized. Two main lattice stripes are identified, and the lattice spacing is 0.297nm and 0.164nm, respectively. The interplanar distances of (110) and (101) crystal planes match (Fig.S3 and Fig.S4).

The Fourier-transform infrared spectroscopy (FTIR) indicates that different plasma treatments did not significantly alter the functional groups of the $ZrO_2$ NPs themselves (Fig 3c). Both $ZrO_2$ NPs and 3Y-TZP NPs samples prepared under different treatment conditions exhibit broad absorption peaks in the range of 3200 to 3500 $cm^{-1}$. The main functional groups of $ZrO_2$ NPs include an absorption peak at approximately 1384 $cm^{-1}$, corresponding to the stretching vibration of Zr-O bonds, and a characteristic absorption peak at approximately 834 $cm^{-1}$ for monoclinic-phase zirconia (m-$ZrO_2$). Due to the vibration of Zr-O and Zr-O-Zr groups, a vibration band is formed in the region of 685-875 $cm^{-1}$, confirming the formation of cubic zirconia, which corresponds to the crystalline structure of zirconia. 3Y-TZP NPs exhibit an absorption peak at 1556 $cm^{-1}$, caused by the vibration of Zr-OH bonds, which is formed by the bending and stretching vibrations of water molecules containing bound water in the sample.

It can be observed from Fig. 3d that when $ZrO_2$ NPs and 3Y-TZP NPs are dispersed in F-12K culture medium solution, their average particle sizes are (234.87 ± 3.38) nm and (290.00 ± 13.30) nm, with polydispersity indices of (0.292 ± 0.108) and (0.362 ± 0.087), respectively. As shown in Fig 3b, the main particle size distribution of Nano-$ZrO_2$ is in the range of 25.0~275.0 nm, while that of 3Y-TZP NPs is in the range of 175.0~325.0 nm. The size distribution results are based on calculations using the NNLS (Non-Negative Least Squares) distribution model. Both zirconia nanoparticles exhibit soft agglomeration in the F-12K culture medium solution. This may be attributed to the high ionic strength and neutral nature, where the presence of salt partially neutralizes the surface charge and reduces electrostatic repulsion. Consequently, zirconia nanoparticles tend to aggregate in the F-12K culture medium solution. The particle size and distribution of nanoparticles can change when they are present in solution.

The results indicate that for $ZrO_2$ NPs, when water is used as the solvent, the nanoparticles exhibit an overall positive charge without plasma treatment (Fig. 3e). However, after plasma treatment for 10, 20, 30, 60, and 120 s, the nanoparticles exhibit an overall negative charge, indicating that the nanoparticle surfaces become negatively charged. $ZrO_2$ NPs tend to aggregate more after plasma treatment, resulting in decreased stability compared to the control group. For 3Y-TZP NPs, when water is used as the solvent, the nanoparticles exhibit an overall negative charge without plasma treatment. After plasma treatment for 10, 20, 30, and 60 s, the nanoparticles exhibit an overall positive charge, indicating that the nanoparticle surfaces become positively charged. However, after plasma treatment for 120 s, the nanoparticle surfaces become negatively charged. Similar to $ZrO_2$ NPs, 3Y-TZP NPs tend to aggregate more after plasma treatment, resulting in decreased stability compared to the control group. The Zeta potential is less than 30 mV in both the control group and the plasma treatment, indicating poor stability of $ZrO_2$ NPs in deionized water.

Fig. 4a illustrates the status of A549 cells after treatment followed by Calcein/PI staining. After treatment with plasma and the addition of $ZrO_2$ NPs to the culture medium, most cells exhibited morphological changes characterized by cellular shrinkage. Microscopic observation revealed phenomena that living cells have clear cell boundaries, whereas dead cells exhibit ruptured cell membranes, blurred cell boundaries, and leakage of cell contents (Fig. S5). From Fig. 4b, it can be observed that the average cell viability of the 50 μL and 100 μL groups after 120 s of plasma treatment is approximately 43.71% and 60.16%, respectively. The 50 μL group exhibits better cytotoxicity compared to the 100μL group, with smaller errors observed in each group. As indicated by Fig. S6, compared to the control group without plasma treatment, the average cell viability after 4 h of culture following 120 s of plasma treatment is approximately 55.11%, while after 24 h of culture, it decreases to around 41.85%. The difference in cell viability between the two culture times is approximately 13.26%. This underscores the significance of investigating the combined use of plasma and $ZrO_2$ NPs. The results suggest that with increasing plasma treatment time, cell viability gradually decreases. Moreover, appropriately extending the post-plasma treatment cell culture time also contributes to a certain extent to reducing cell activity.

As the concentration of $ZrO_2$ NPs increased from 50 to 300 μg/mL, its cytotoxicity towards A549 cells also increased. When the concentration reached 300 μg/mL, the cell viability was (31.37±3.72) % and (35.68±3.23) % for cell densities of 10,000 cells and 20,000 cells, respectively (Fig. 4c and Fig. S6). At a cell density of 30,000 cells, the highest cytotoxicity was observed at a concentration of 250 μg/mL, with a cell viability of (34.18±3.22) %. When the $ZrO_2$ NPs concentration exceeded 100 μg/mL, all groups showed a significant difference compared to the control group at a significance level of 0.01%. As shown in Fig. 5c and Fig. S6, 3Y-TZP NPs exhibit no cytotoxicity towards A549 cells within the range of 50 to 300 μg/mL. When the concentration of 3Y-TZP NPs is 300 μg/mL, the proliferative effect on A549 cells is most pronounced.

The reason for the result could be attributed to the different effects and impacts of nanoparticles with different shapes on cells. Compared to $ZrO_2$ NPs, 3Y-TZP NPs have an additional composition of 3% mol $Y_2O_3$. Rod-shaped nano $Y_2O_3$ has a proliferative effect on cells. Additionally, 3Y-TZP is the zirconia material with the highest mechanical properties at room temperature, characterized by high strength, high fracture toughness, high wear resistance, heat resistance, and corrosion resistance, which makes it non-toxic to A549 cells.[42] Moreover, it can promote the proliferation of A549 cells within a certain concentration range. We observed that at a cell count of 10,000, the cell viability is (121.27±7.28)% at a 3Y-TZP NPs concentration of 50 μg/mL, indicating a proliferative effect on cells.

$ZrO_2$ NPs at a concentration of 300 μg/mL were added for a 4h incubation period, followed by exposure to plasma for 120 s and subsequent incubation (Fig. 4d). The cell viability rate was 27.34±3.41%. The results showed significantly lower cell viability compared to using plasma or Nano-$ZrO_2$ alone, and there was no significant difference between the two combination methods statistically. For 3Y-TZP NPs at a concentration of 300 μg/mL followed by exposure to plasma for 120 seconds, the cell viability rates were 71.44±4.48%. The results indicate that plasma and $ZrO_2$ NPs induce cell death in A549 cells, while 3Y-TZP NPs can promote cell proliferation to some extent (Fig. 4d and Fig. S6).

Figs. 5a-b and Fig. S7 demonstrate significant damage caused by $ZrO_2$ NPs to the cellular mitochondria. During apoptosis, a loss of mitochondrial membrane potential may occur. In contrast, Nano-3Y-TZP does not cause significant damage to the mitochondria. Comparing the fluorescence intensity ratio, it is evident that as the concentration of $ZrO_2$ NPs increases, the fluorescence intensity decreases, ultimately reaching 0.61±0.07. In addition, we observed that the fluorescence intensity decreased

as the plasma treatment time extended. After 120s of plasma treatment, the fluorescence intensity decreased to 0.72±0.02 compared to the control group. This indicates that plasma treatment induces mitochondrial depolarization in a time-dependent manner.

Fig. 5d indicates that the intracellular ATP concentration significantly decreased in the plasma-treated group and $ZrO_2$ NPs group, while it significantly increased in the 3Y-TZP NPs group. Since mitochondria are the primary energy suppliers within cells and can generate ATP required for cellular functions,[43] abnormalities or damage to mitochondria following plasma and $ZrO_2$ NPs treatments may inhibit ATP synthesis, leading to cellular energy deficiency and affecting normal cellular metabolism and survival. CAP, $ZrO_2$ NPs, and 3Y-TZP NPS treatment all lead to an increase in intracellular ROS and RNS. The results indicate that the combination of plasma and $ZrO_2$ NPs induces the highest increase in intracellular ROS and RNS. Exposure to nanoparticles may lead to an increase in intracellular inflammatory responses, as nanoparticles can activate inflammation-related signaling pathways.[44, 45] These inflammatory factors further stimulate oxidative stress and the generation of ROS and RNS (Fig. 5c, 5e, and 5f).

Fig. S8 demonstrates a positive correlation between the concentrations of RNS and ROS with plasma treatment time, suggesting a potential relationship between plasma exposure duration and its toxic effects on cells. When only $ZrO_2$ NPs were added to the culture medium, no generation of ROS and RNS was detected in the culture medium. However, once $ZrO_2$ NPs entered the interior of the cells, they may react with oxygen molecules inside the cells, producing peroxides and free radicals, among other reactive oxygen species. These reactive oxygen species can damage molecular structures such as cell membranes, proteins, and DNA, leading to cell death.[46-48] 3Y-TZP NPs at a certain level can promote A549 cell proliferation. When only 3Y-TZP NPs were added to the culture medium, the concentrations of RNS and ROS in the medium gradually increased with increasing 3Y-TZP NPs concentration. Within the concentration range, RNS concentrations range from 5uM to 35uM, while ROS concentrations range from 5uM to 60uM, showing a concentration-dependent trend. Moderate levels of ROS and RNS can regulate cellular metabolic activity, especially processes related to energy metabolism and nucleic acid synthesis.[49] They can modulate mitochondrial function and activity of the mitochondrial respiratory chain, promoting ATP synthesis and energy supply, thus providing necessary energy for cell proliferation.

Fig. 6a shows the principal component analysis results of the intra-group repeatability of the samples. Overall, the differences between the control group, CAP group, 3Y-TZP NPs group, and CAP@3Y-TZP NPs group are small, and the inter-group correlation is small. The $ZrO_2$ NPs group and CAP@$ZrO_2$ NPs group are significantly different from the control group. Fig. S9 shows the intersection and statistics of differentially expressed genes in cells between groups. Figs. 6b and 6c show the differentially expressed genes ($q<0.05$, $|log_2FC|>1$) through comparative analysis. After CAP treatment, MAGEA10, GJB5, GPR87, LY6D, and GAS5-AS1 indicated UP expression, which belonged to the tumor antigen gene family or genes related to tumors. SPNS2, ASGR2, ERICH4, and SLC17A4 also appeared DOWN expression. These encode metabolic enzymes are related to cell metabolic processes. It is suggested that the activity of some enzymes will be affected after plasma treatment, which is related to abnormal regulation and metabolic changes in the development of cancer cells. We also found that some genes involved in glucose metabolism and energy production processes had significant changes, such as CSNK1E, CYP1A1, SPNS2, CPA4, KLHL30, JUN, PGM5, PLPPR1, SLC13A2, and SLC17A4. It indicates plasma treatment affects the energy of cells, metabolic and glycolytic pathways. In addition, some genes associated with ferroptosis, such as ACSL5, CP, CTH, NOX1, SAT1, TF, and TP53, showed changes to a certain extent. The GO enrichment results in Fig. 6d show that iron ion transport and oxygen binding were significantly enriched, revealing the relevant mechanism of ferroptosis in A549 cells caused by plasma treatment.

We also compared the genes between the CAP@$ZrO_2$ NPs group and the control group (Fig. 6d). We found that the above-mentioned genes had changed to varying degrees. The combined use of CAP and $ZrO_2$ NPs led to changes in the CTSW gene, indicating the cell lysosomal function and protein degradation ability enhanced. Cancer cells usually show a high dependence on glycolytic metabolism (Warburg effect). By overexpressing HK2, glucose uptake and metabolism can be increased to meet the energy needs of their rapid growth and proliferation. As a key enzyme of glucose metabolism, the HK2 gene plays a vital role in maintaining in cell energy metabolism balance and survival. After the combined treatment of CAP and $ZrO_2$ NPs, this gene showed a significant down-regulation trend, suggesting that they interfere with the metabolic pathways of tumor cells by inhibiting HK2 activity or reducing its expression level, inhibiting tumor growth and spread. Fig. S10 shows the corresponding differential expression volcano plots, including $ZrO_2$-vs-Ctrl, CAP@$ZrO_2$-vs-CAP, CAP@$ZrO_2$-vs-$ZrO_2$, 3Y-TZP-vs-Ctrl, CAP@3Y-TZP-vs-Ctrl、CAP@3Y-TZP-vs-3Y-TZP, and CAP@3Y-TZP-vs-CAP.

The various ROS and RNS generated by CAP in the culture medium can lead to an increase in intracellular ROS levels.[50-54] Plasma treatment can generate a series of reactive species such as free radicals, hydrogen peroxide, and nitric oxide. These reactive species are highly active and can undergo redox reactions with molecules inside the cell, leading to the generation of reactive oxygen and nitrogen species.[55-57] Furthermore, nanoparticles themselves may undergo chemical reactions such as electron transfer and free radical generation, which can result in the production of ROS and RNS within the cells.[58, 59] These species can then be released into the cell, triggering oxidative stress reactions.

We performed GO (Fig. S11), KEGG Pathway (Fig. S12), Reactome (Fig. S13), and WikiPathways (Fig. S14) enrichment analysis based on hypergeometric distribution algorithm for differentially expressed genes. The potential pathway mechanism diagram of plasma and $ZrO_2$ NPs respectively acting on A549 cells was drawn as shown in Fig. 7a. The enrichment analysis results revealed that plasma inhibits PI3K /Akt signaling pathway, inhibiting PI3K activity, reducing PIP3 production, and decreasing Akt phosphorylation. It weakens the proliferation ability of cancer cells and induces apoptosis via regulating downstream molecules of the PI3K/Akt pathway (such as mTOR, GSK-3, Bad, and Cyclin D1).[60, 61] In addition, the results show that plasma activates the FoxO signaling pathway. FoxO transcription factors can regulate the expression of cell cycle-related genes and inhibit the proliferation of tumor cells by inhibiting cell cycle progression. Meanwhile, it can regulate the metabolic pathways of lung cancer cells, such as glucose metabolism and lipid metabolism, thereby affecting the survival of lung cancer cells by regulating metabolic pathways. FoxO transcription factors can regulate the expression of apoptosis-related genes, enhance the sensitivity of tumor cells to apoptotic signals, and promote apoptosis of cancer cells.[62, 63] They can also regulate the expression of genes related to cell migration and metastasis, such as metastasis-related protein (MET), matrix metalloproteinases (MMPs), etc., thereby reducing the infiltration and metastasis of lung cancer cells. A549 cells after $ZrO_2$ NPs resulted into TGF-β pathway enriched. $ZrO_2$ NPs trigger an intracellular signaling cascade, leading to the activation of TGF-β receptors and oxidative phosphorylation of downstream signaling molecules, thereby activating TGF-β pathway.[64] In addition, $ZrO_2$ NPs activate apoptosis-related signaling pathways, such as the mitochondrial pathway, caspase signaling pathway, and pathways regulated by Bcl-2 family proteins. The activation of these signaling pathways can trigger the apoptosis program and lead to cell death.

Fig. 7b shows the potential pathway mechanism of the combined application of CAP and $ZrO_2$ NPs, and suggests that the combination of the two may achieve synergistic effects. It is found that compared with individual treatments, the gene enrichment effect on the TGF-β pathway is more significant. It indicates that the combined effect has a higher aggregation effect and further enhances the therapeutic effect.[65, 66] The TGF-β pathway mainly regulates apoptosis-related gene expression (Smad2/3) and induces cells to enter a proliferation arrest state, leading to apoptosis of cancer cells and limiting their ability to proliferate.[67] The TGF-β/Smad pathway is a key signaling pathway within the TGF-β superfamily of cytokines. It is responsible for transmitting the TGF-β signal from the cell surface to the nucleus, where it regulates gene expression. TGF-β can induce cell cycle arrest and apoptosis by activating its downstream Smad signaling pathway, especially in the case of abnormal cell proliferation. Since Smad2 and Smad3 are the main effector molecules in the TGF-β signaling pathway, their downregulation may lead to weakened TGF-β signaling, thereby reducing the proliferation of tumor cells. Here, q-PCR results showed that CAP@ZrO2 induced TGF-β, Smad2, Smad3, Smad4, and Bcl-2 expression were downregulated, while Bax expression was upregulated (Fig. S18). In addition, combined effects are related to the TNF signaling pathway, Wnt signaling pathway, Sphingolipid signaling pathway, Hippo signaling pathway, and Rap-1 signaling pathway. Furthermore, we employed GSEA software to conduct gene analysis (Fig. S15), and the results showed that both the CAP group and the $ZrO_2$ NPs group were enriched at the top of mitochondrial translation compared with the control group, revealing changes in mitochondrial function. The abnormalities in mitochondrial protein synthesis and translation processes related to mitochondrial dysfunction, energy metabolism, disorders, and other processes.[68, 69] And this change has the most significant effect when the two act together.

Through analysis of the enrichment results of the 3Y-TZP NPs group, it was found that the ErbB signaling pathway was activated. ErbB activates downstream signaling molecules, such as RAS-MAPK and PI3K-AKT pathways, to promote cytoskeletal reorganization, epithelial-to-mesenchymal transition, and basement membrane degradation processes, which thereby enhance the migration and invasion capabilities of cancer cells.[70] The ErbB signaling pathway also can promote cell cycle progression, enhance cell survival signals, and inhibit apoptosis, thereby giving cancer cells an advantage in proliferation and survival.[71, 72] 3Y-TZP NPs treatment resulted in enrichment at the top of cell division-related pathways, suggesting strong proliferation activity (Fig. S16). In addition, by activating downstream signaling molecules, such as VEGF (vascular endothelial growth factor) and FGF (fibroblast growth factor), the ErbB signaling pathway can promote the formation of new blood vessels, provide oxygen and nutrients to tumors, and further promote tumor growth and transfer.[73, 74]

For the synergistic effect of CAP and 3Y-TZP NPs (CAP@3Y-TZP), the combination enriched the ABC (ATP-binding cassette) transport pathway and Wnt signaling pathway. In the Wnt signaling pathway, CAP@3Y-TZP down-regulated the expression of LRP5 and LRP6 blocking the transmission of Wnt signals, which is similar to the regulatory mechanism of CAP@ $ZrO_2$. It resulted in a decrease in the transcription levels of lung cancer cell proliferation-related genes, inhibiting cell cycle progression and thereby inhibiting cell proliferation. Down-regulating the expression of LRP5 and LRP6 may also lead to the loss of cell polarity and reduce migration ability, thereby affecting processes such as cytoskeletal rearrangement, cell adhesion, and extracellular matrix degradation.[75] In addition, there are other potential pathway mechanisms, including the Hippo signaling pathway, cGMP-PKG signaling pathway, and MAPK signaling pathway. By analyzing the key genes of the Hippo signaling pathway, it was found that CAP@3Y-TZP inhibited the Hippo signaling pathway, which is completely opposite to CAP@ $ZrO_2$. In general, some pathways pointed to cell apoptosis, and some pathways pointed to cell proliferation or inhibited apoptosis in CAP@3Y-TZP. The potential mechanisms of CAP@3Y-TZP signaling pathways are shown in Fig. S17.

## CONCLUSIONS

CAP combined with $ZrO_2$ NPs offers a novel strategy for lung cancer treatment. This combination therapy affects mitochondrial function in lung cancer cells, inducing endoplasmic reticulum stress and DNA damage. We have found that the combination of CAP and $ZrO_2$ nanoparticles can induce lung cancer cell apoptosis by activating the TGF-β pathway. Additionally, RNA sequencing results have provided multiple possible mechanisms for the killing of lung cancer cells by CAP alone and combination therapy. When CAP acts alone on cells, it may induce ferroptosis through the lipid peroxidation pathway or induce apoptosis through oxidative stress. Furthermore, the combined treatment can also induce apoptosis or necrosis through TNF, WNT, or Hippo signaling pathways, offering direction for subsequent mechanistic research.

**ACKNOWLEDGEMENTS**

This work was supported by the National Key Research and Development Program of China (2022YFE0126000, Z.C.), the Guangdong Basic and Applied Basic Research Foundation (2022A1515011129, Z.C.), Shenzhen Science and Technology Major Project (KJZD20230923114406014, Z.C.), Shenzhen Medical Academy of Research and Translation (C2301011, Z.C.), Science and Technology Innovation Commission of Shenzhen (JCYJ20190809100813289, R.Z.), and Shenzhen Clinical Research Center for Oral Diseases (20210617170745001-SCRC202201016, Y.W.).

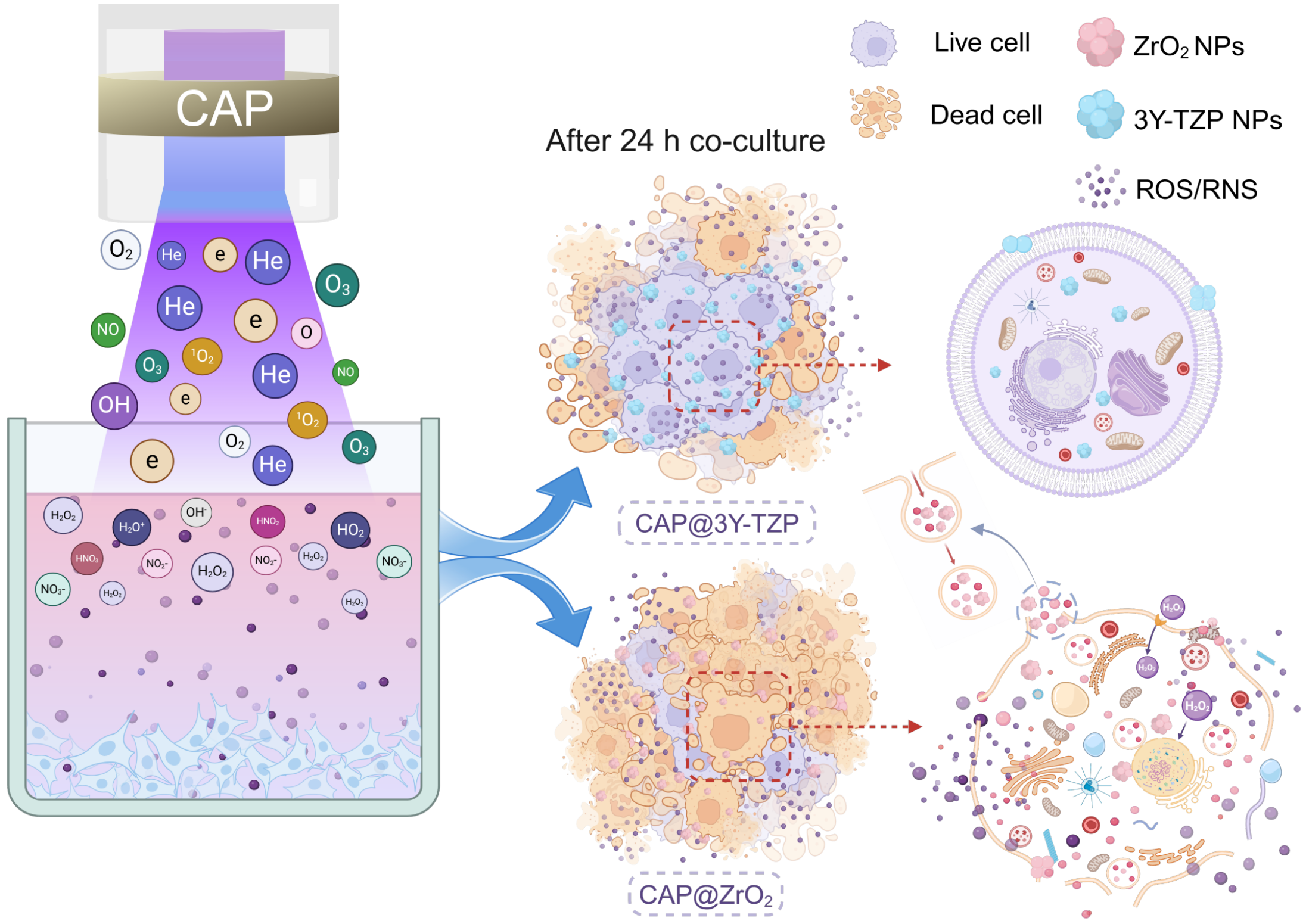

**Fig. 1**. Schematic of the interaction between CAP with $ZrO_2$/3Y-TZP NPs and A549 cancer cells.

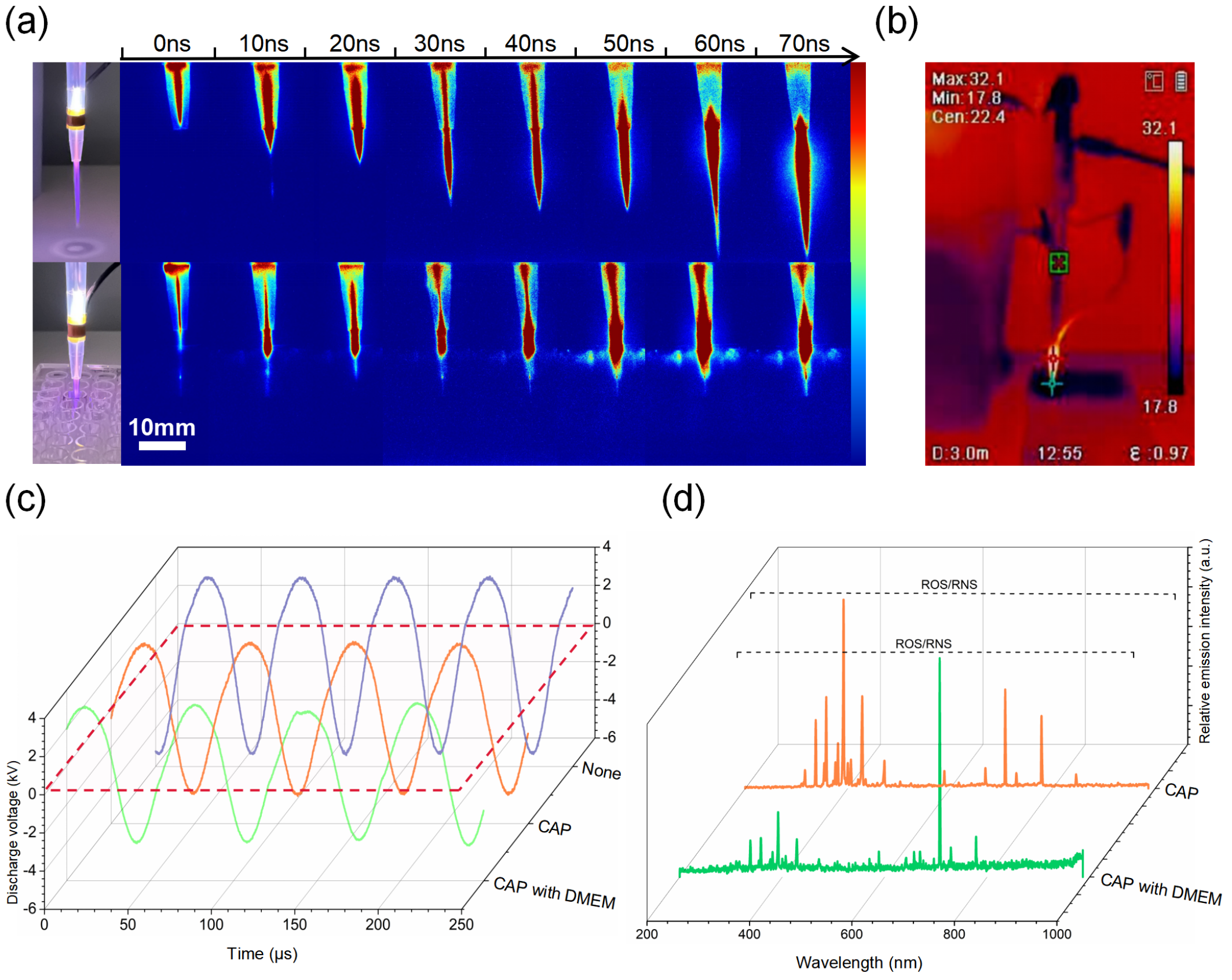

**Fig. 2**. Helium plasma jet device (Fig. S1) and its characteristics. (a) The discharge evolution process of the helium plasma jet and the discharge evolution process acting on the 96-well plate with culture medium; (b) The discharge temperature of the helium plasma; (c) The discharge voltage waveform of the helium plasma; (d) Optical emission spectrum of the plasma jet.

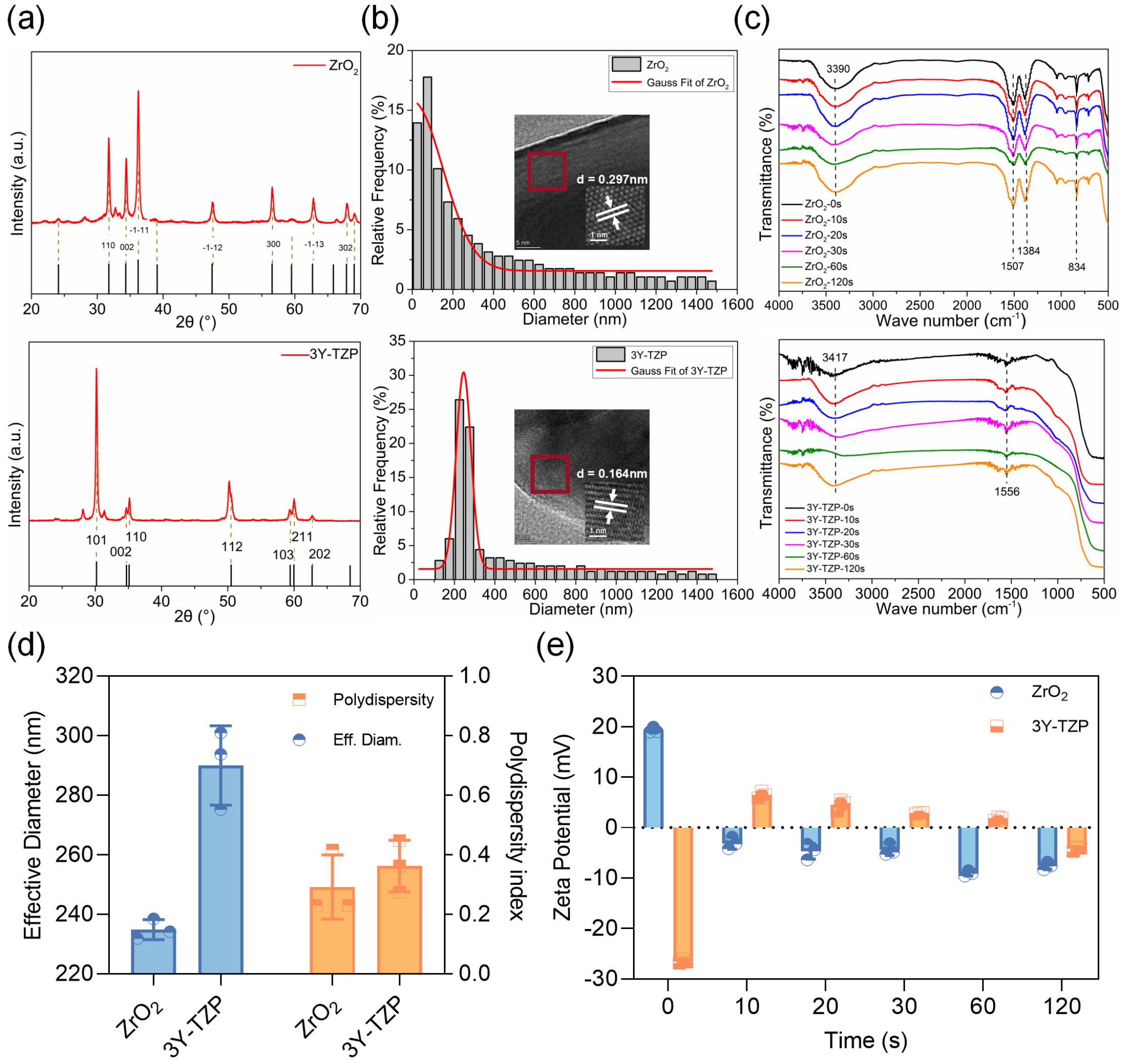

**Fig. 3**. Structural characterization of $ZrO_2$ NPs and 3Y-TZP NPs. (a) X-ray diffraction pattern; (b) Hydrated particle size distribution and lattice fringe image at 5nm; (c) Fourier-transform infrared spectroscopy (FTIR) spectra after plasma treatment for 0s, 10s, 20s, 30s, 60s, and 120s; (d) Effective hydrated particle size and polydispersity index in culture medium; (e) Zeta potential after plasma treatment for 0s, 10s, 20s, 30s, 60s, and 120s.

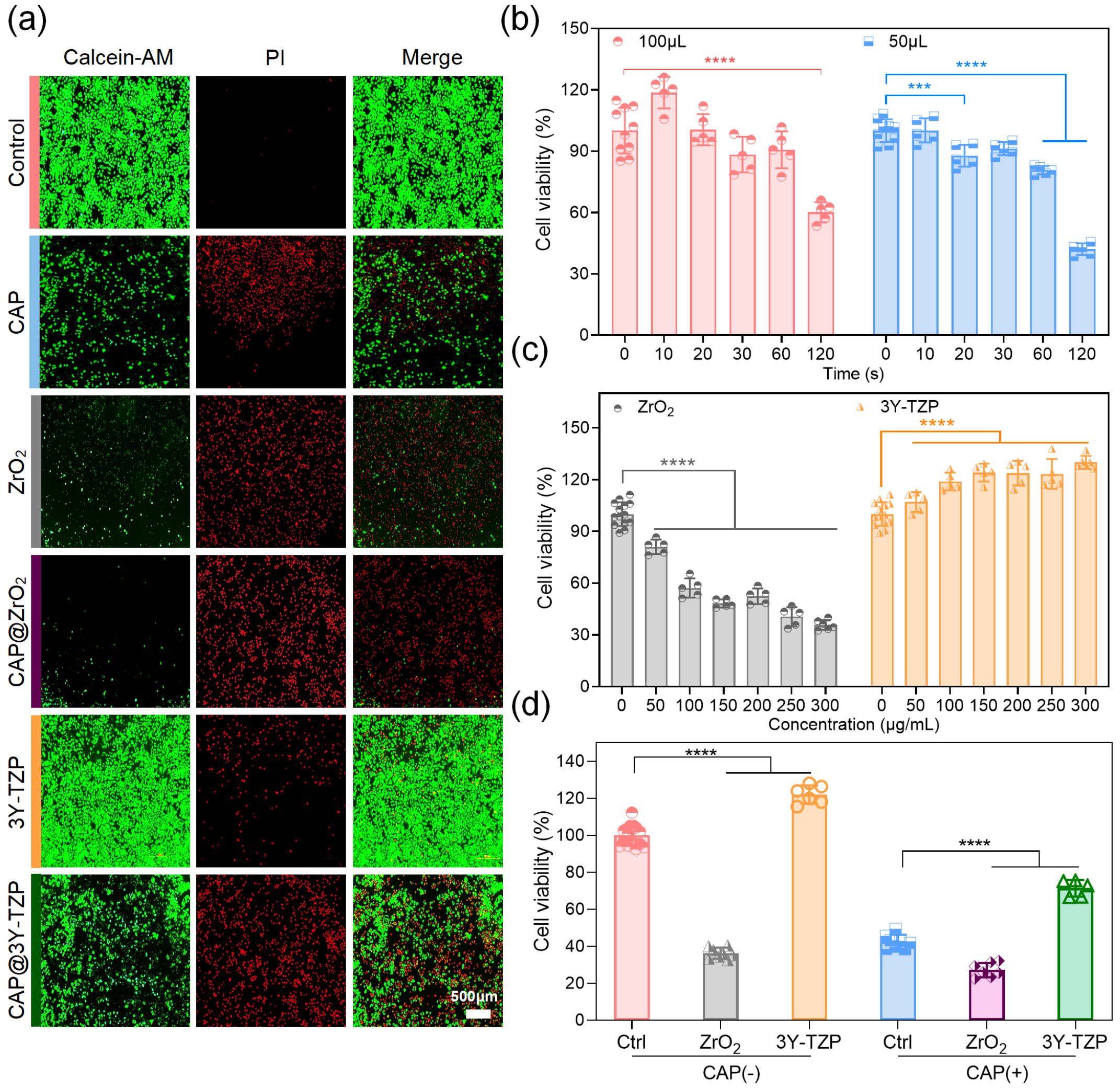

**Fig. 4**. Evaluation of the cytotoxicity of different treatment conditions on A549 cells using Calcein/PI staining and MTT assay. (a) Viability of A549 cells stained with Calcein/PI method, scale bar: 500μm; (b) Impact of different durations and modes of plasma treatment on the viability of A549 cells; (c) Effect of treatment with different concentrations of $ZrO_2$ NPs and 3Y-TZP NPs on the viability of A549 cells after 24 hours of incubation; (d) Impact of individual and combined treatment of $ZrO_2$ NPS and 3Y-TZP NPs with plasma on the viability of A549 cells. (n = 5, *P < 0.05, **P < 0.01, ***P< 0.001, and ****P < 0.0001).

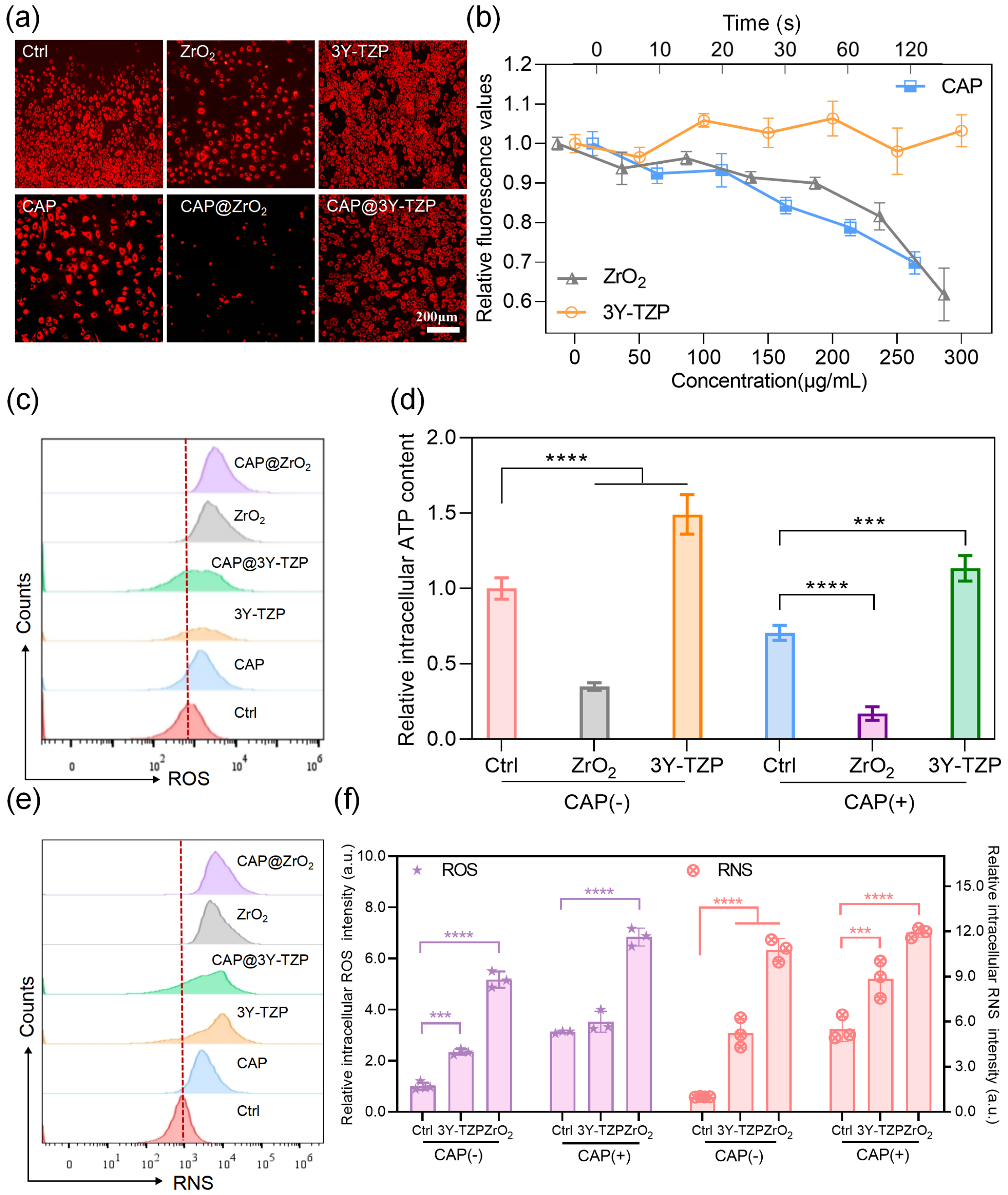

**Fig. 5**. (a) Cellular mitochondrial status observed under an inverted fluorescence microscope after staining with TMRE under different treatment conditions, scale bar: 200μm; (b) Relative fluorescence intensity values of mitochondrial membrane potential in A549 cells under different plasma treatment times and concentrations of $ZrO_2$ NPs or 3Y-TZP NPs; (c and e) Peak values of intracellular ROS and RNS in FITC channel under various treatments of plasma, $ZrO_2$ NPs, or 3Y-TZP NPs alone or in combination; (d) Analysis of the effect of different treatments on ATP concentration changes in A549 cells; (f) Average fluorescence intensity of intracellular ROS and RNS in cells treated with plasma, $ZrO_2$ NPs, or 3Y-TZP NPs alone or in combination. (n = 3, *$P < 0.05$, **$P < 0.01$, ***$P < 0.001$, and ****$P < 0.0001$).

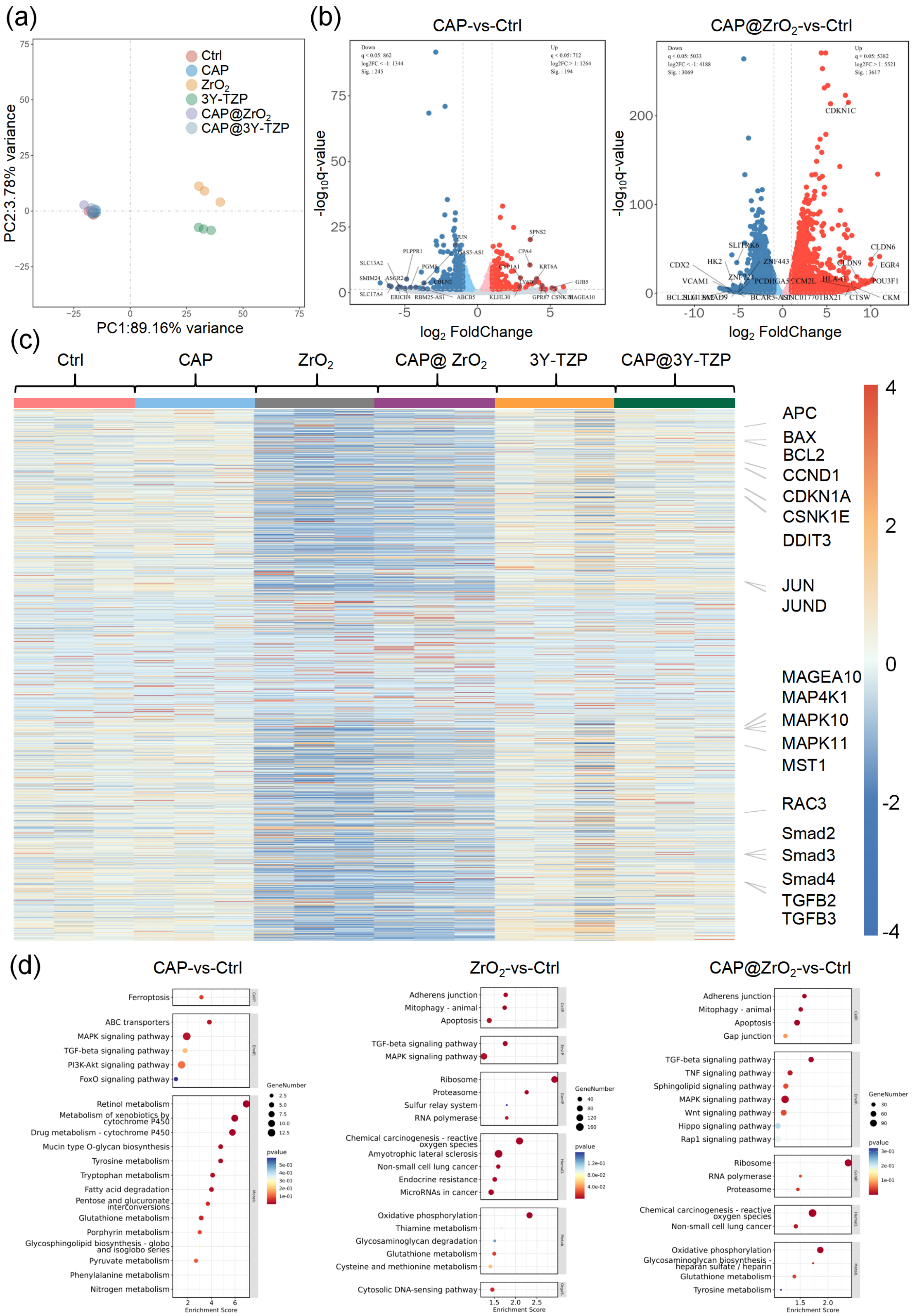

**Fig. 6**. (a) Principal component analysis diagram of Ctrl, CAP, $ZrO_2$ NPs, 3Y-TZP NPs, CAP@$ZrO_2$ and CAP@3Y-TZP; (b) Differential expression volcano diagram corresponding to CAP-vs-Ctrl and CAP@$ZrO_2$-vs-Ctrl. Red represents up-regulated genes, blue represents down-regulated genes; (c)Ctrl, CAP, $ZrO_2$ NPs, 3Y-TZP NPs, CAP@$ZrO_2$ and CAP@3Y-TZP differential gene grouping cluster analysis heat map; (d) The KEGG enriched pathways corresponding to CAP-vs-Ctrl, $ZrO_2$-vs-Ctrl, and CAP@$ZrO_2$-vs-Ctrl.

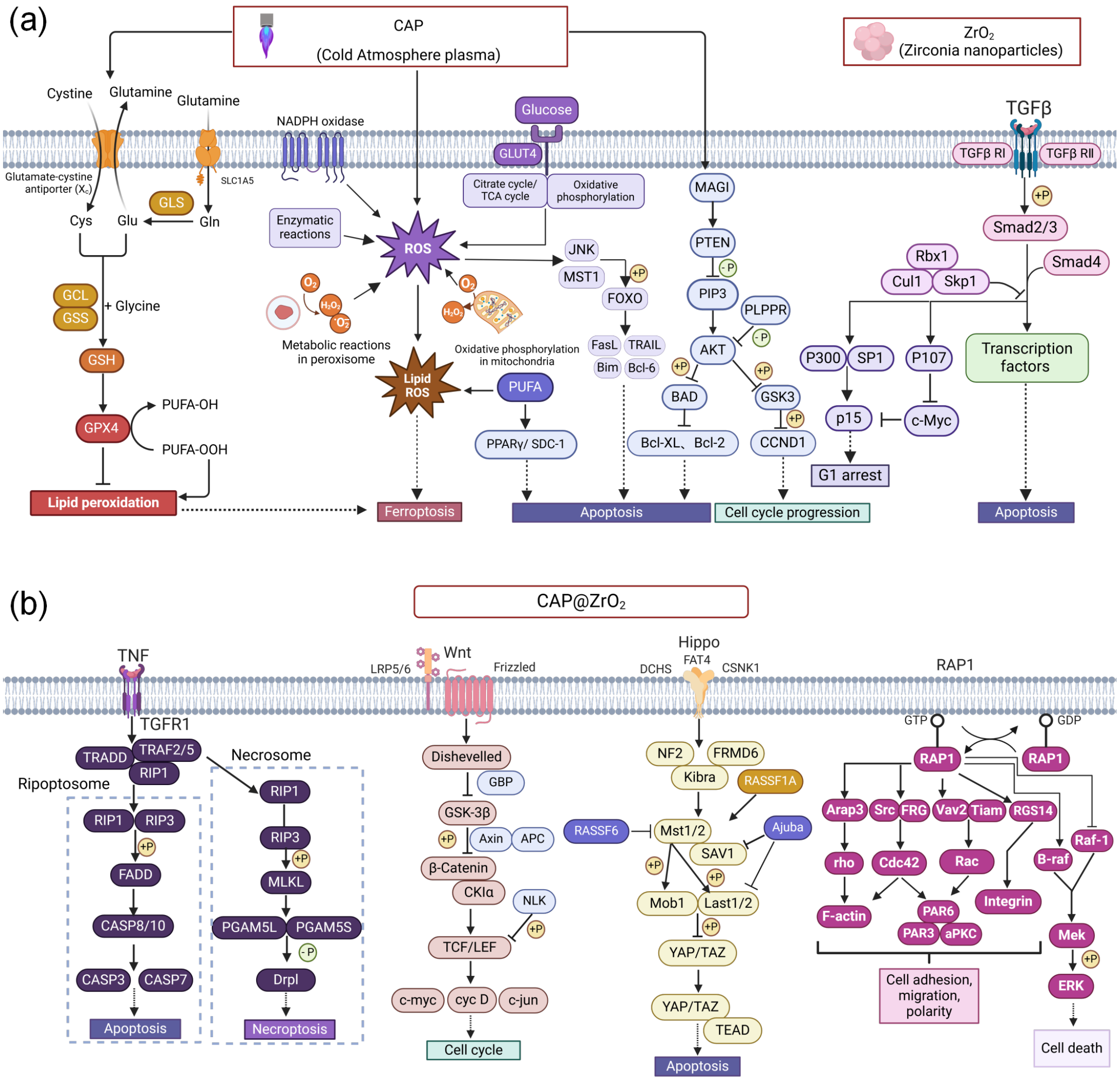

**Fig. 7**. (a) Illustration of the signaling pathway potential mechanisms of CAP and $ZrO_2$ NPs. (b) Illustration of the signaling pathway mechanisms of CAP@ $ZrO_2$.

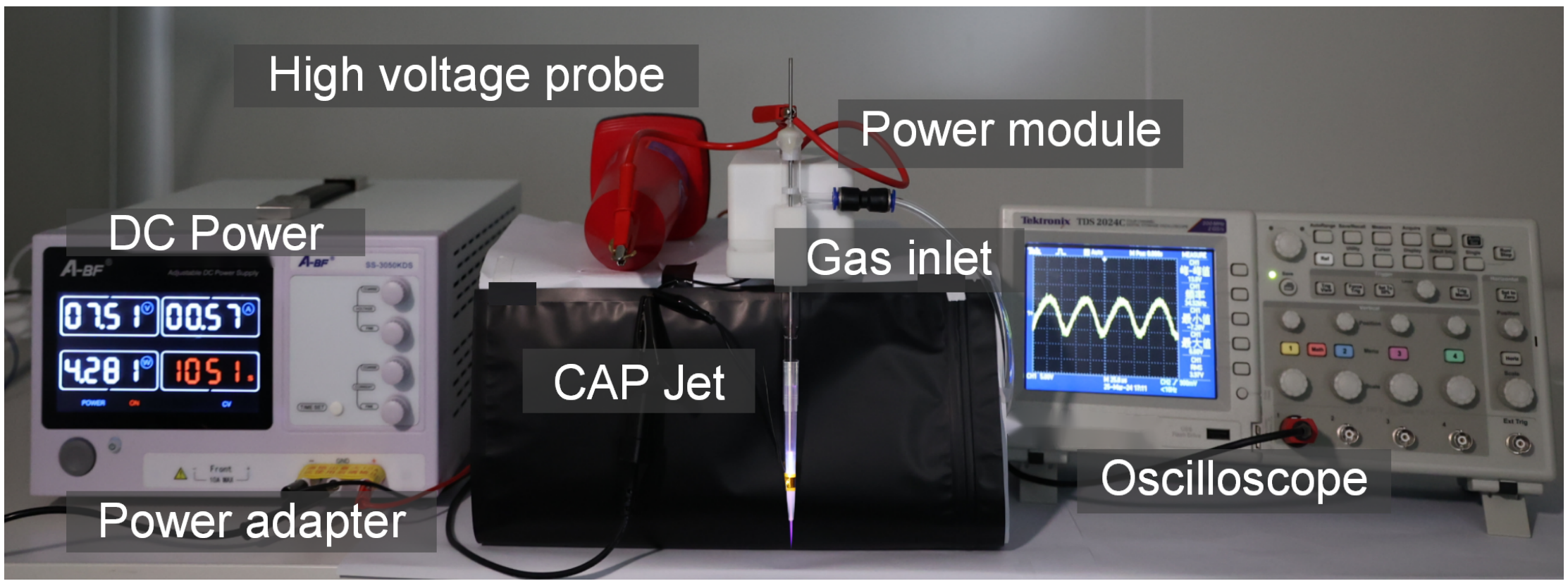

Fig. S1. Schematic diagram of the helium plasma jet device.

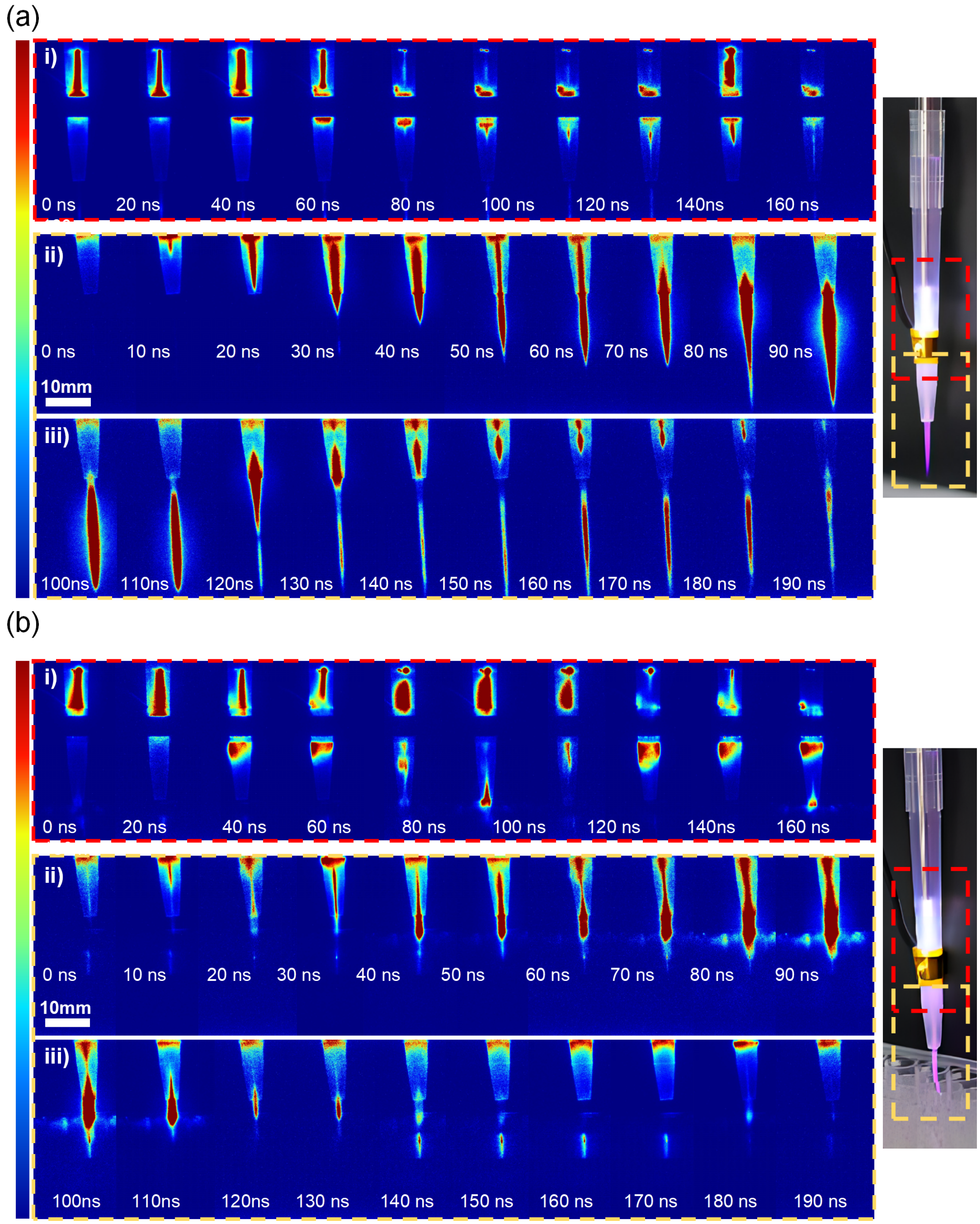

Fig. S1. (a) A time-resolved intensified camera captures and records the discharge process of the helium plasma jet. The row (i) depicts the discharge evolution from the metal rod to the metal ring electrode, while the rows (ii) and (iii) illustrate the plasma discharge evolution from the metal ring electrode to the outlet, encompassing the generation to the annihilation of the plasma. (b) Building upon (a), a 96-well plate containing culture medium is positioned at the outlet of the plasma jet.

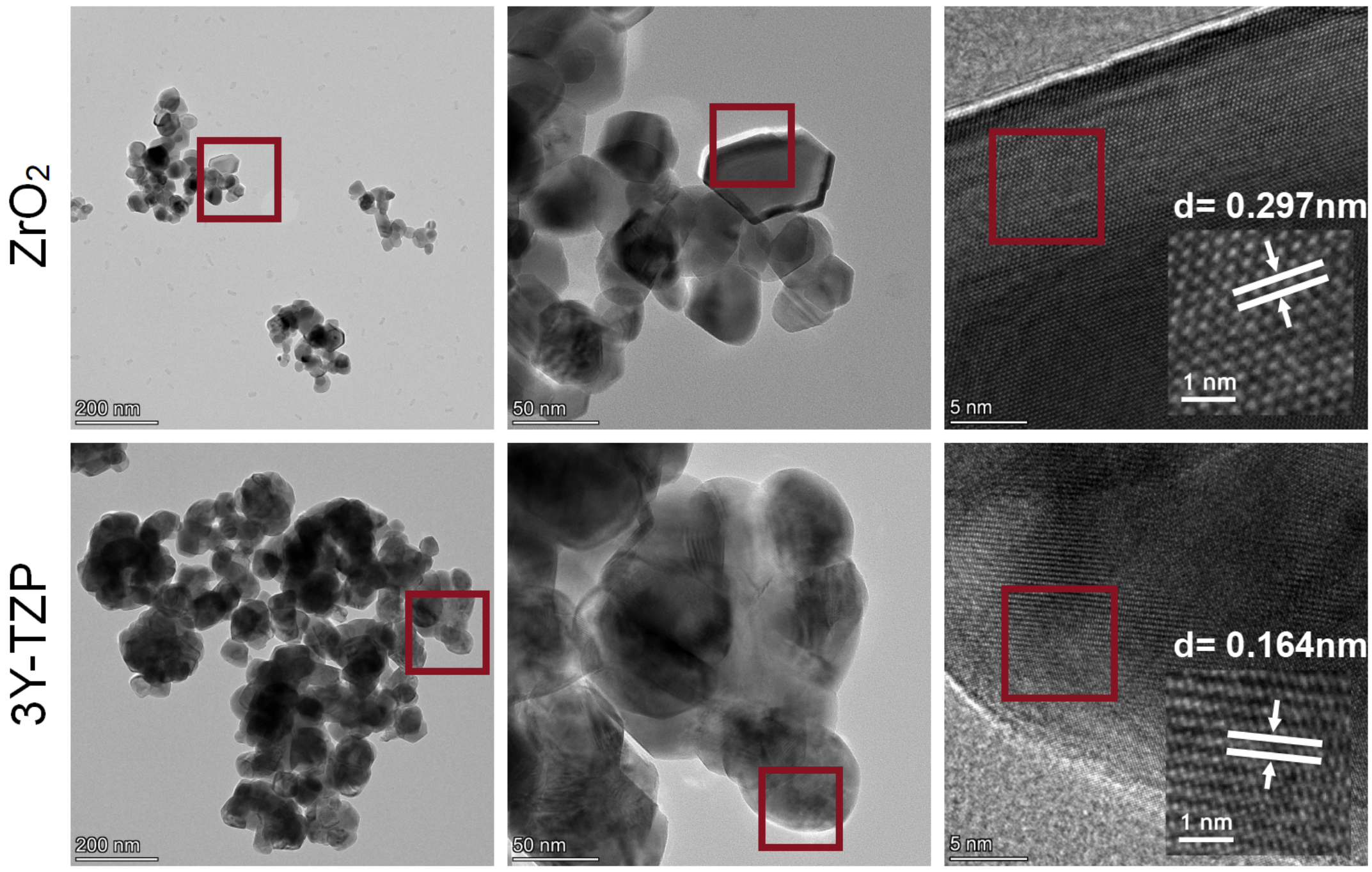

Fig. S2. TEM images at the scales of 200nm and 50nm, as well as lattice fringe images at a scale of 5nm.

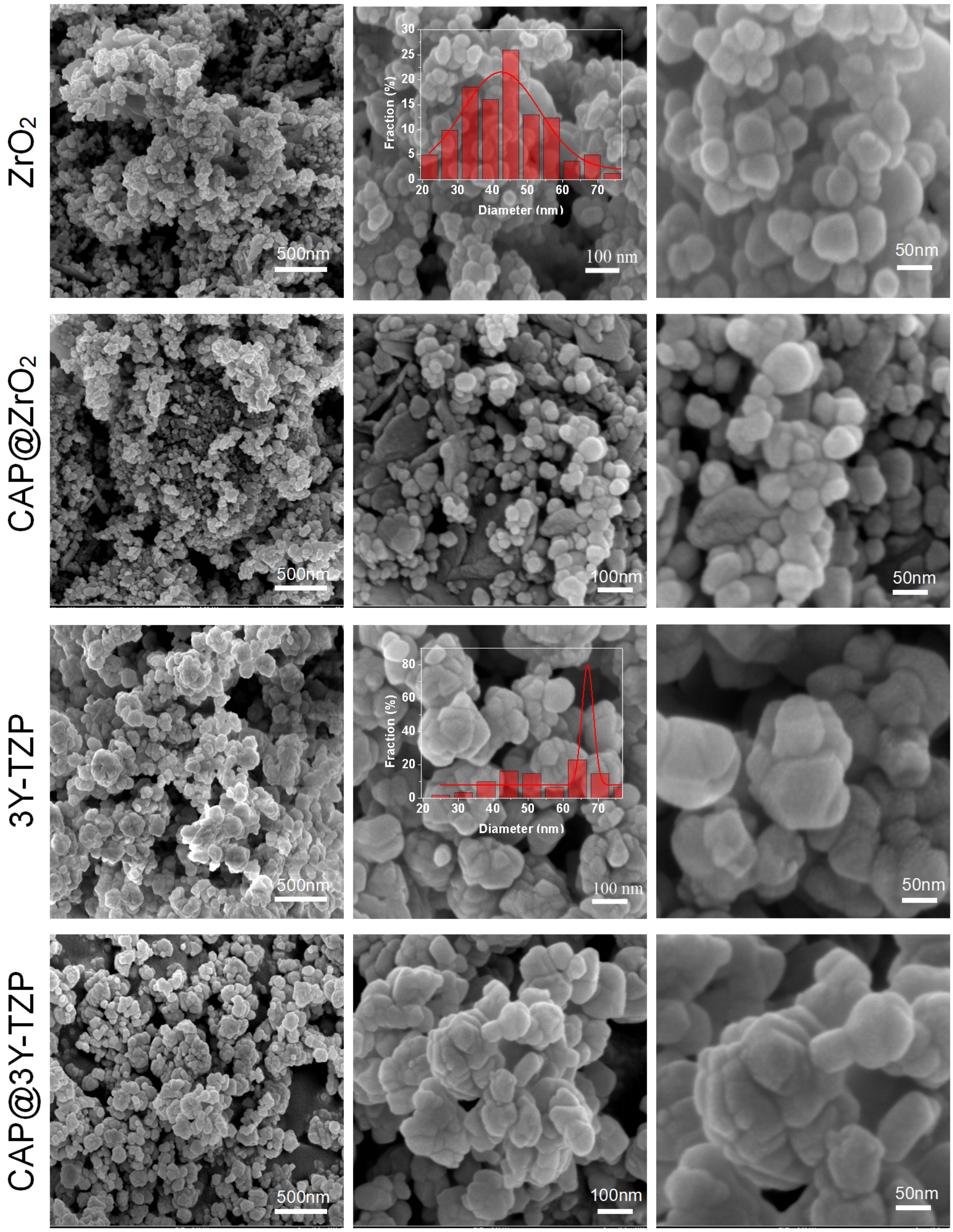

Fig. S3. SEM images at scales of 50nm, 100nm, and 500nm and its particle size distribution curve at a scale of 100nm;

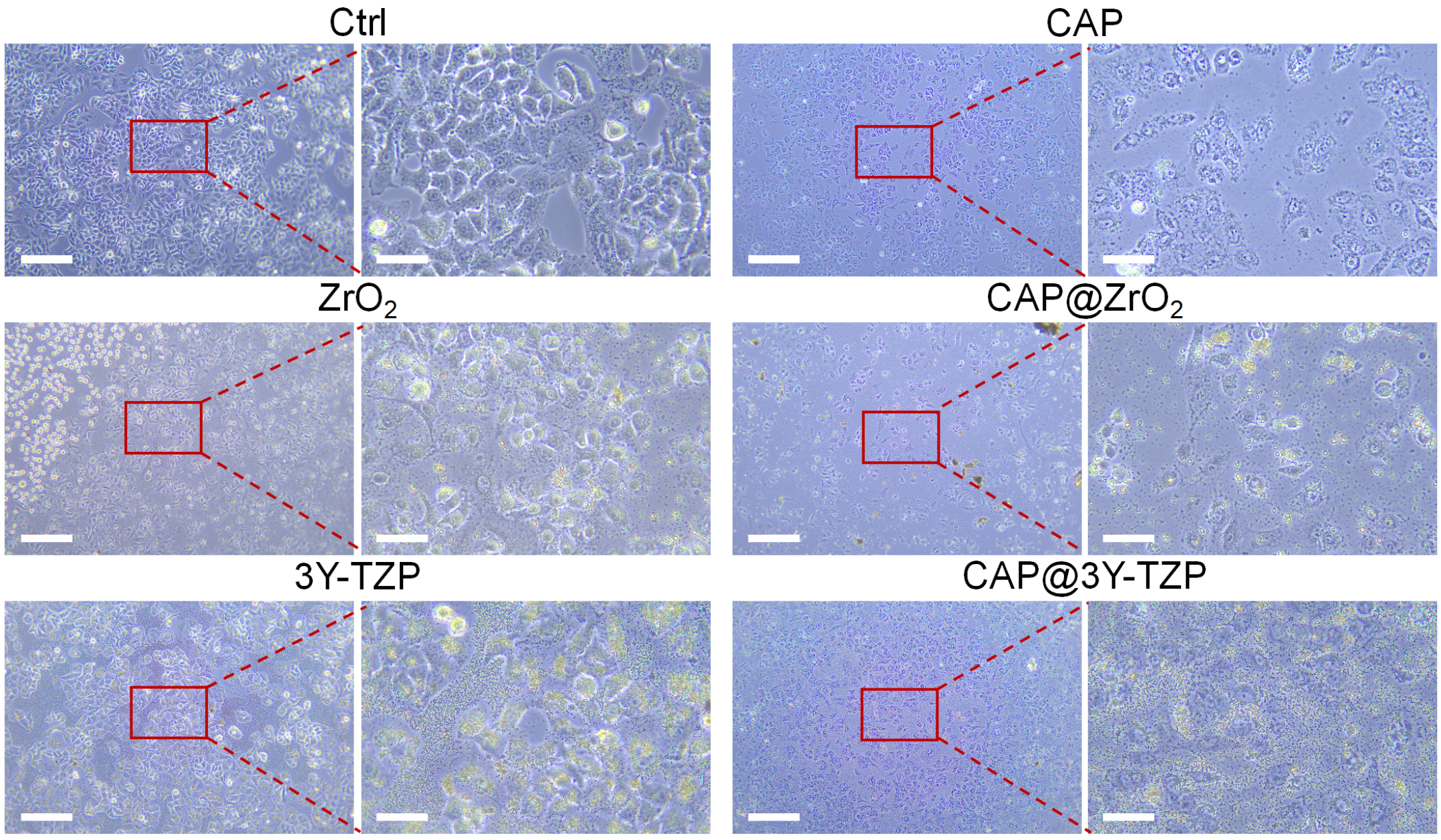

Fig. S5. Observation of cell states under different treatment conditions using an inverted microscope. The magnification levels are 10x and 40x, with a scale of 100μm.

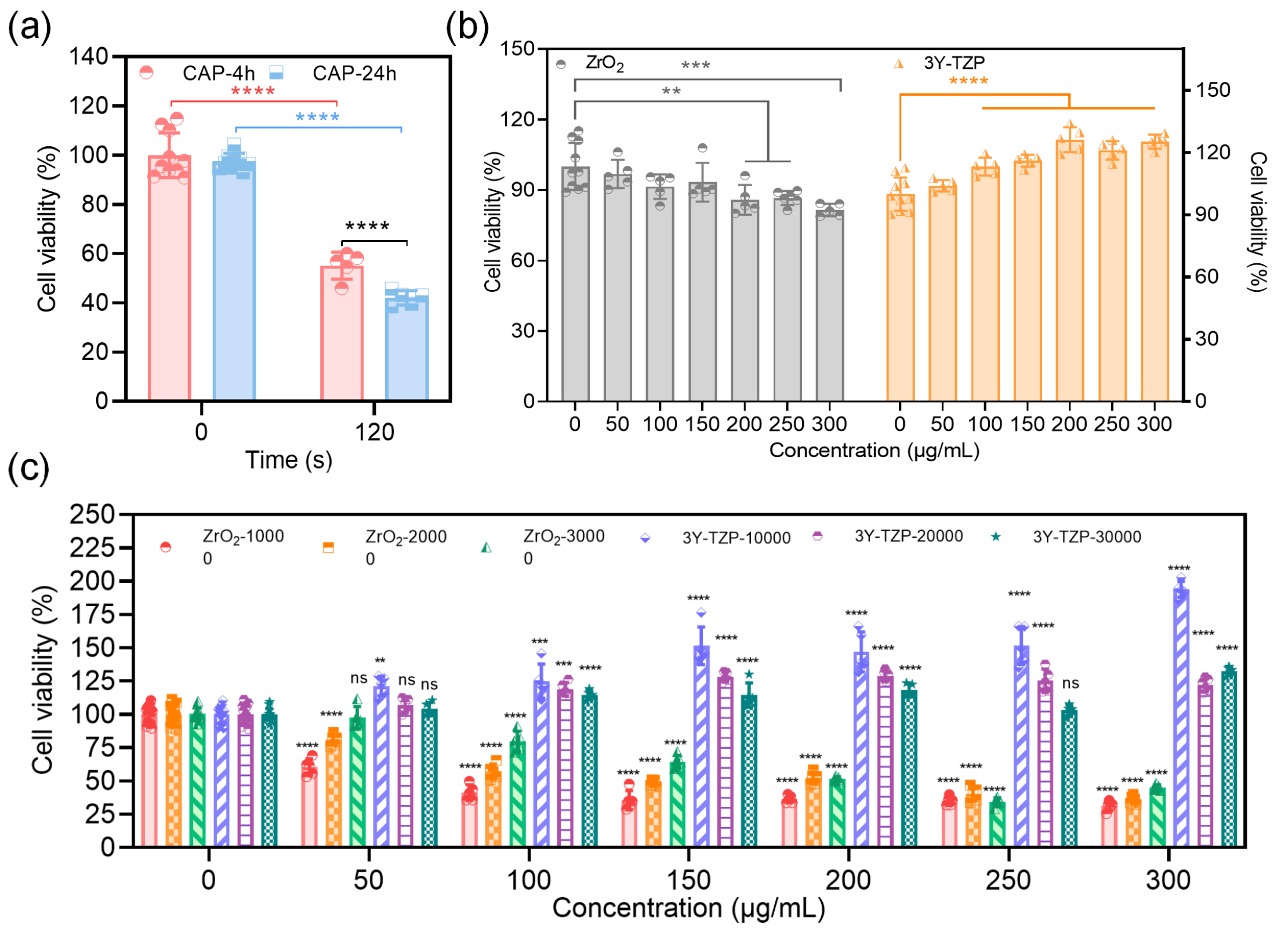

Fig. S6. (a) The effect of continued cultivation for 4h and 24h after 120 seconds of plasma treatment on the viability of A549 cells. (b) The effect of different concentrations of $ZrO_2$ NPs and 3Y-TZP NPs on the viability of A549 cells after 4 hours of cultivation. (c) The effect of different cell densities and concentrations of $ZrO_2$ NPs and 3Y-TZP NPs on the viability of A549 cells.

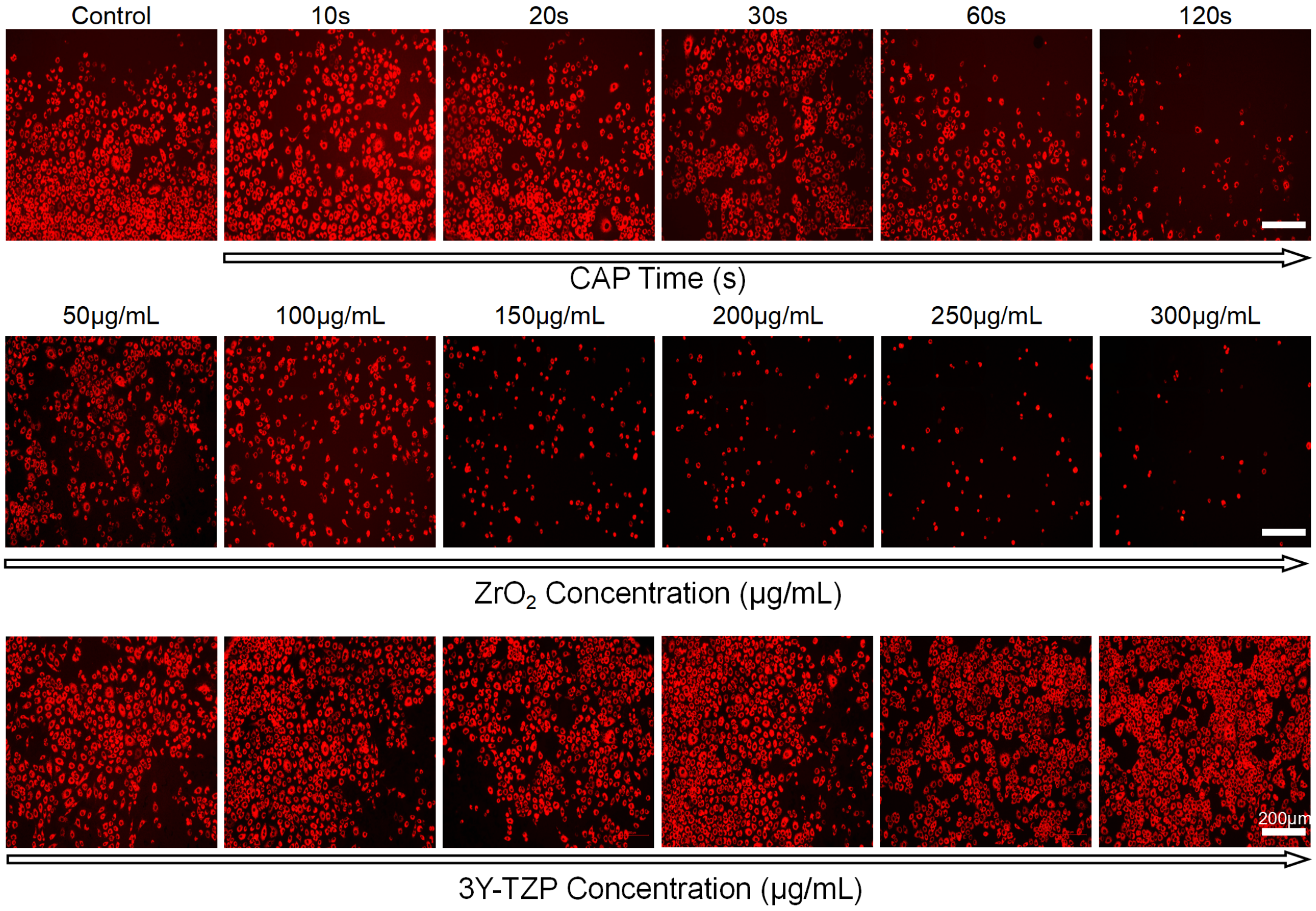

Fig. S7. Observation of cell mitochondrial status using an automated inverted fluorescence microscope after staining with TMRE. The scale bar is set to 200µm.

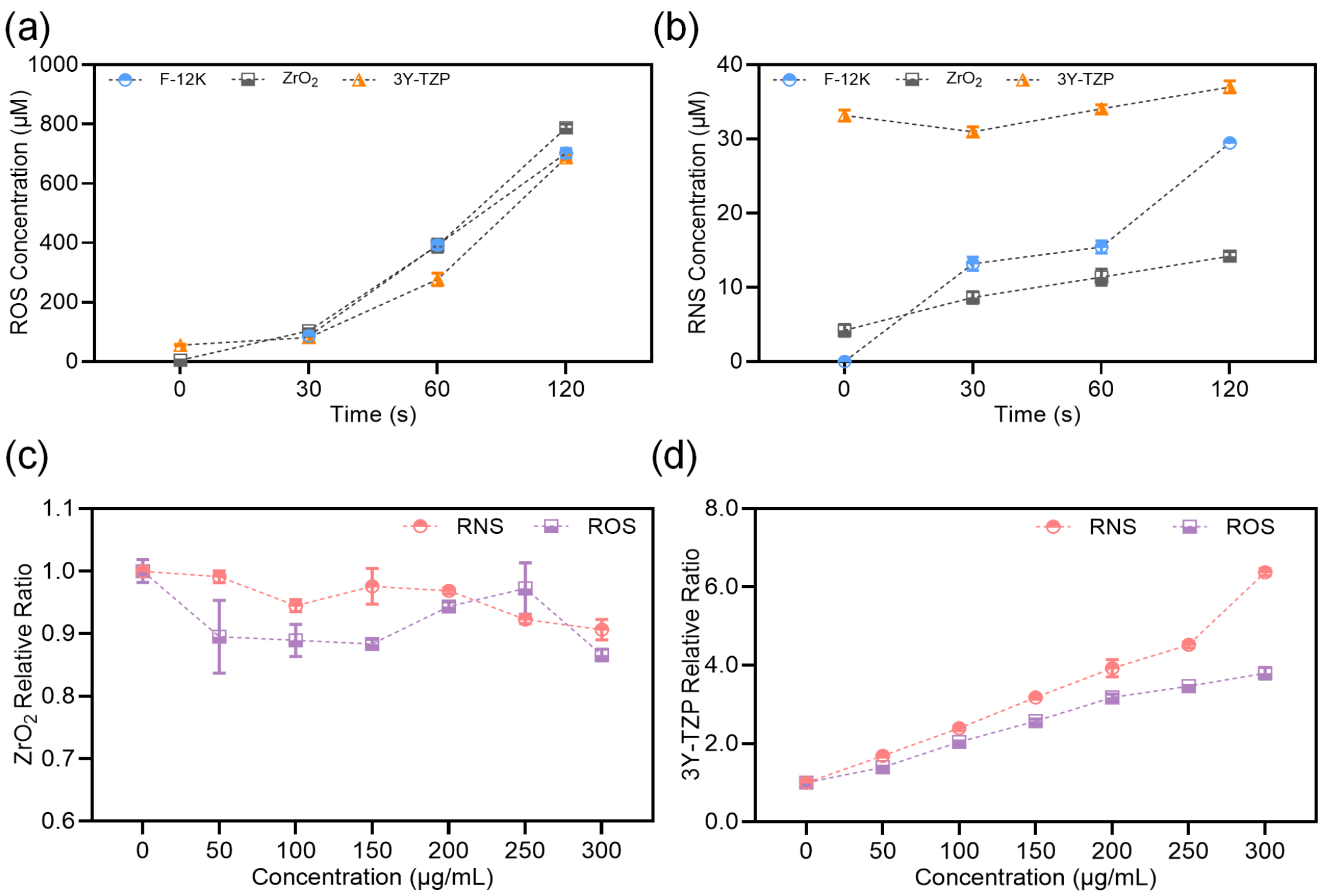

Fig. S4. (a, b) Concentrations of reactive RNS and ROS in F-12K culture medium, $ZrO_2$ NPs suspension, and 3Y-TZP NPs suspension after plasma treatment for 30, 60, and 120 seconds; (c, d) Relative ratios of RNS and ROS in $ZrO_2$ NPs and 3Y-TZP NPs.

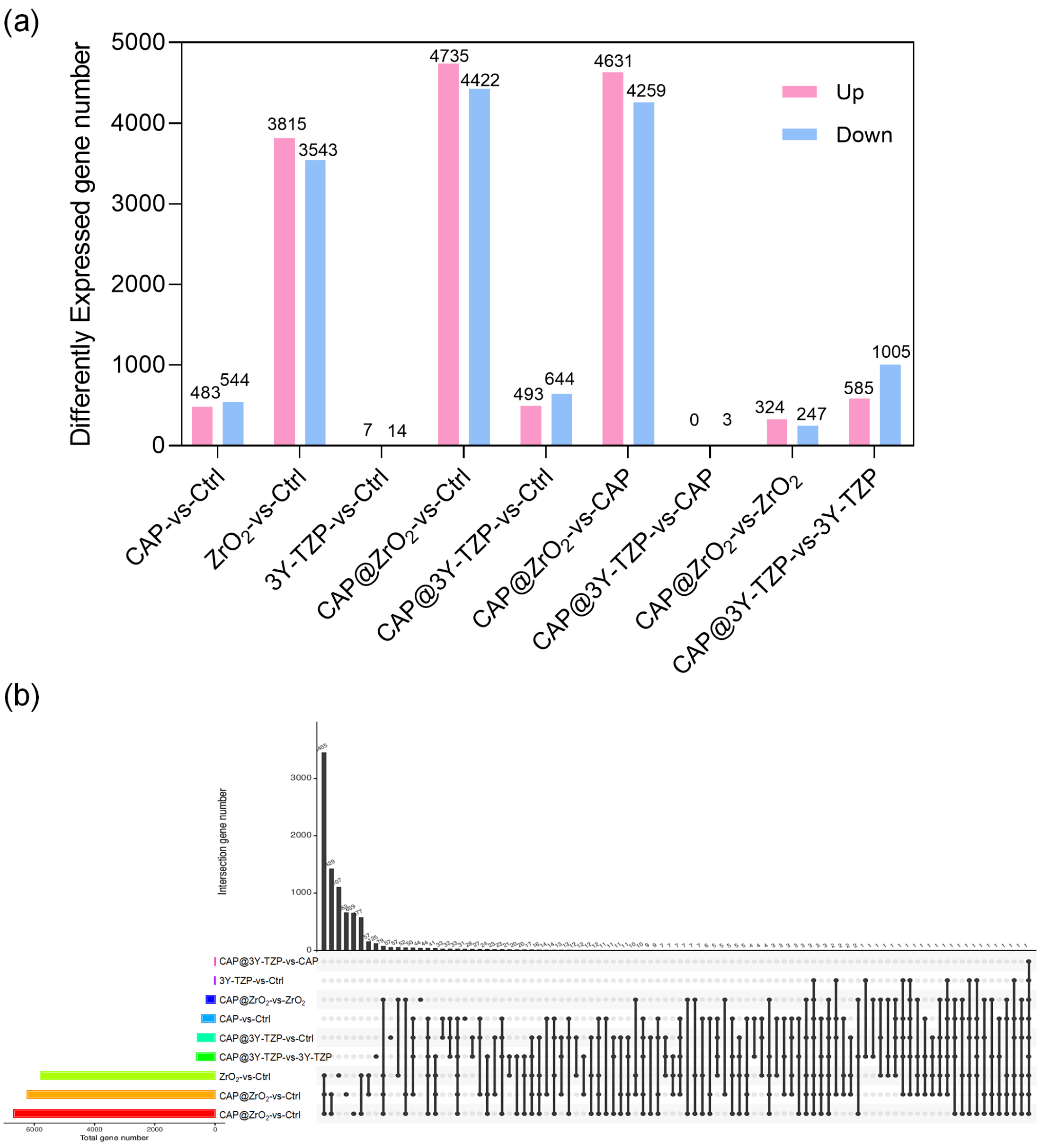

Fig. S5. (a) Statistical diagram of up-regulated genes and down-regulated genes in different groups; (b) common and unique differentially expressed genes between different comparison groups.

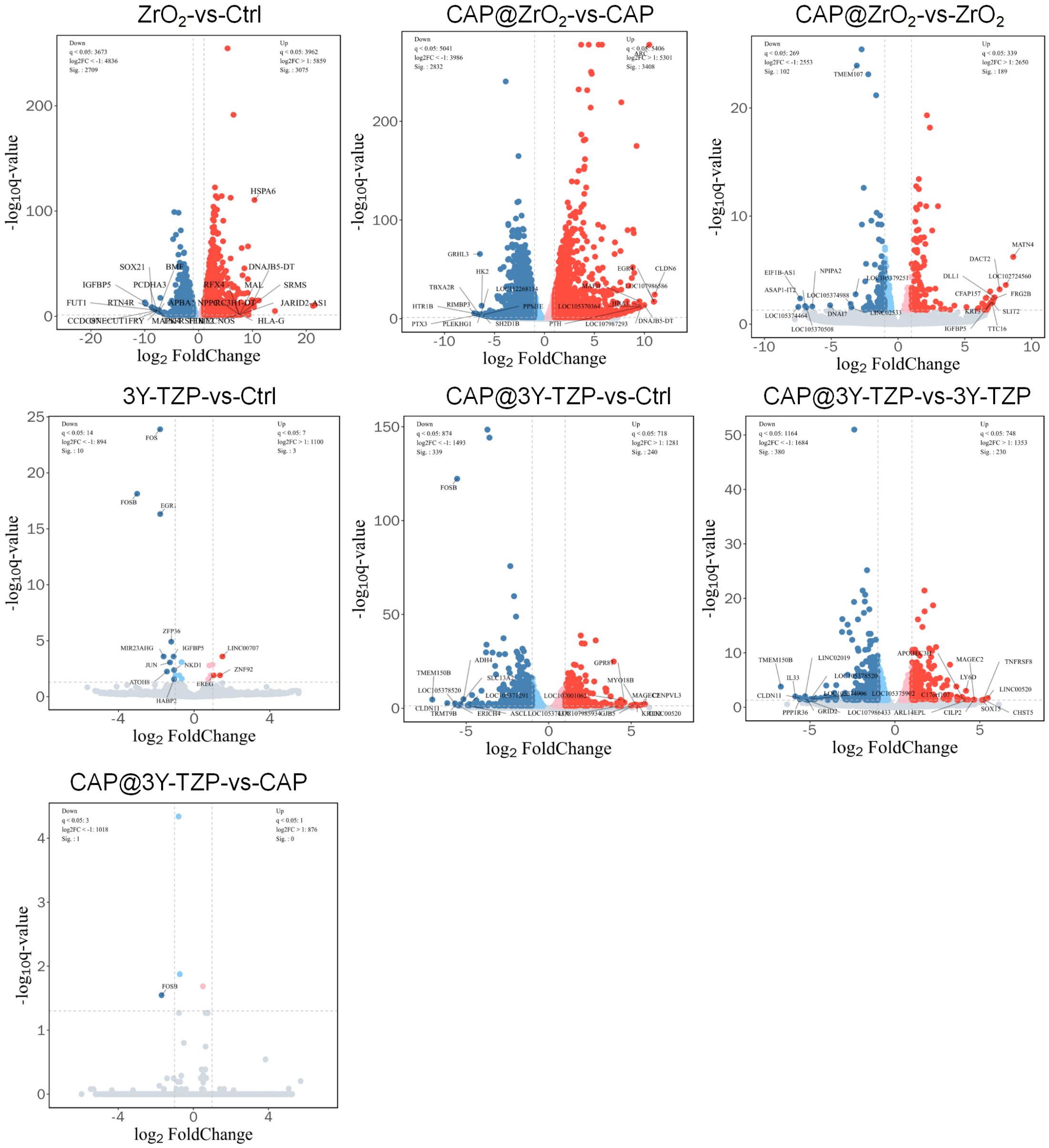

Fig.S6. Volcano plots illustrating the differentially expressed genes for the following comparisons: $ZrO_2$-vs-Ctrl, CAP@$ZrO_2$-vs-CAP, CAP@$ZrO_2$-vs-$ZrO_2$, 3Y-TZP-vs-Ctrl, CAP@3Y-TZP-vs-Ctrl, CAP@3Y-TZP-vs-CAP, and CAP@3Y-TZP-vs-3Y-TZP. In these plots, red color represents upregulated genes, while blue color represents downregulated genes.

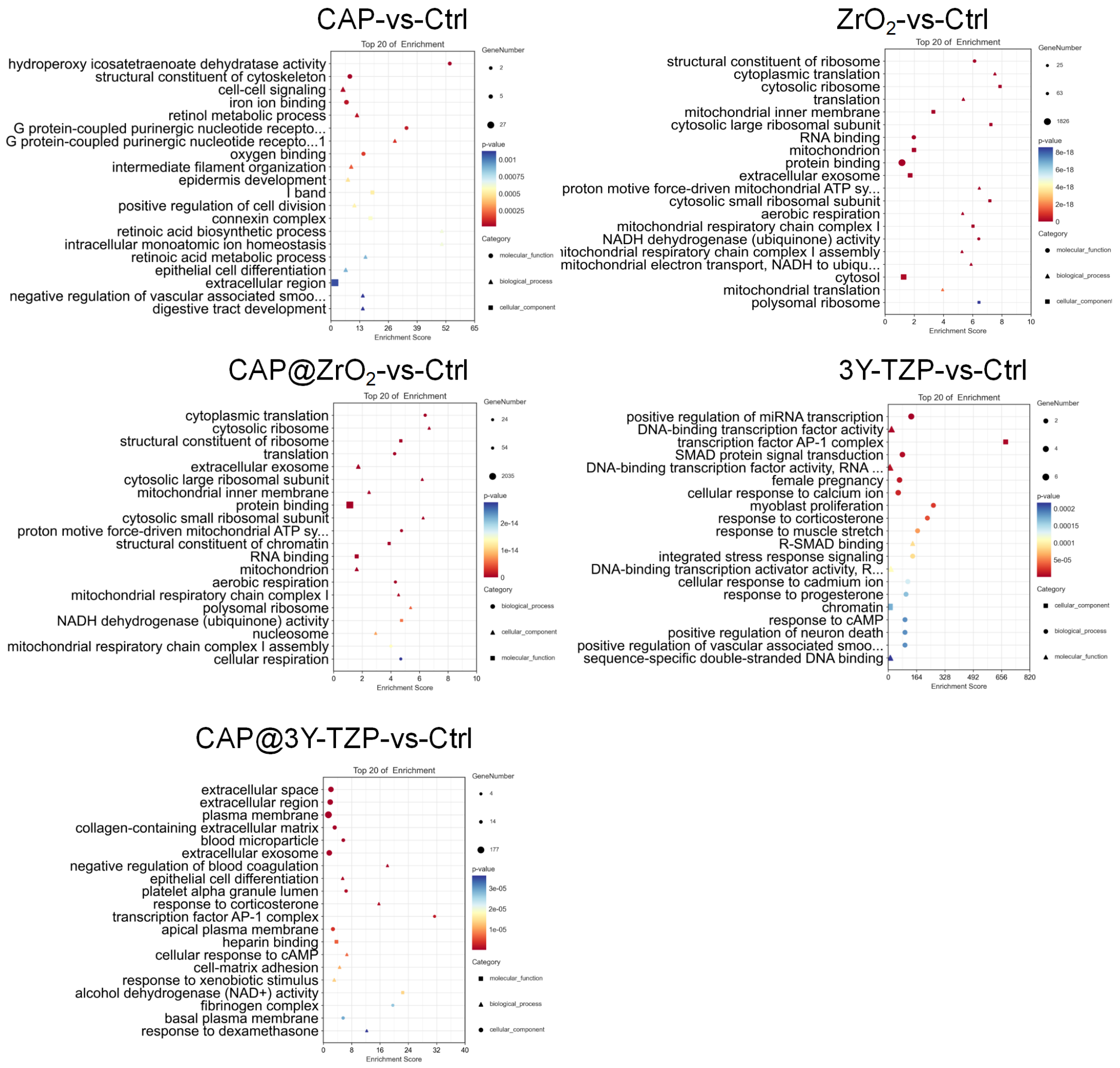

Fig. S7. GO enrichment TOP20 bubble chart between different groups: CAP-vs-Ctrl, $ZrO_2$-vs-Ctrl, CAP@$ZrO_2$-vs-Ctrl, 3Y-TZP-vs-Ctrl, and CAP@3Y-TZP-vs-Ctrl.

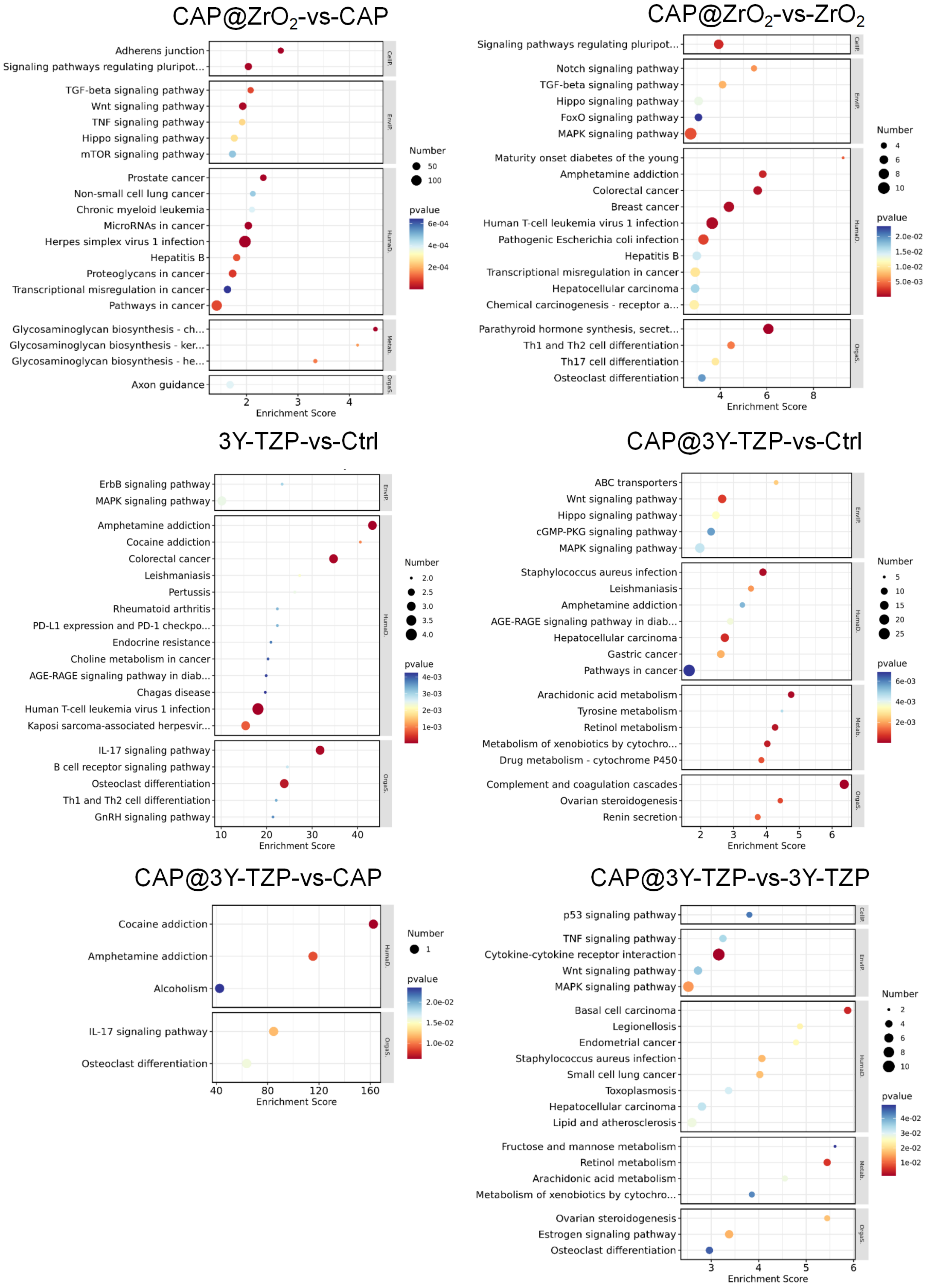

Fig. S8. Corresponding KEGG enriched pathways compared between different groups top20 bubble chart: CAP@$ZrO_2$-vs-CAP, CAP@$ZrO_2$-vs-$ZrO_2$, 3Y-TZP-vs-Ctrl, CAP@3Y-TZP-vs-Ctrl, CAP@3Y-TZP-vs-CAP, and CAP@3Y-TZP-vs-3Y-TZP.

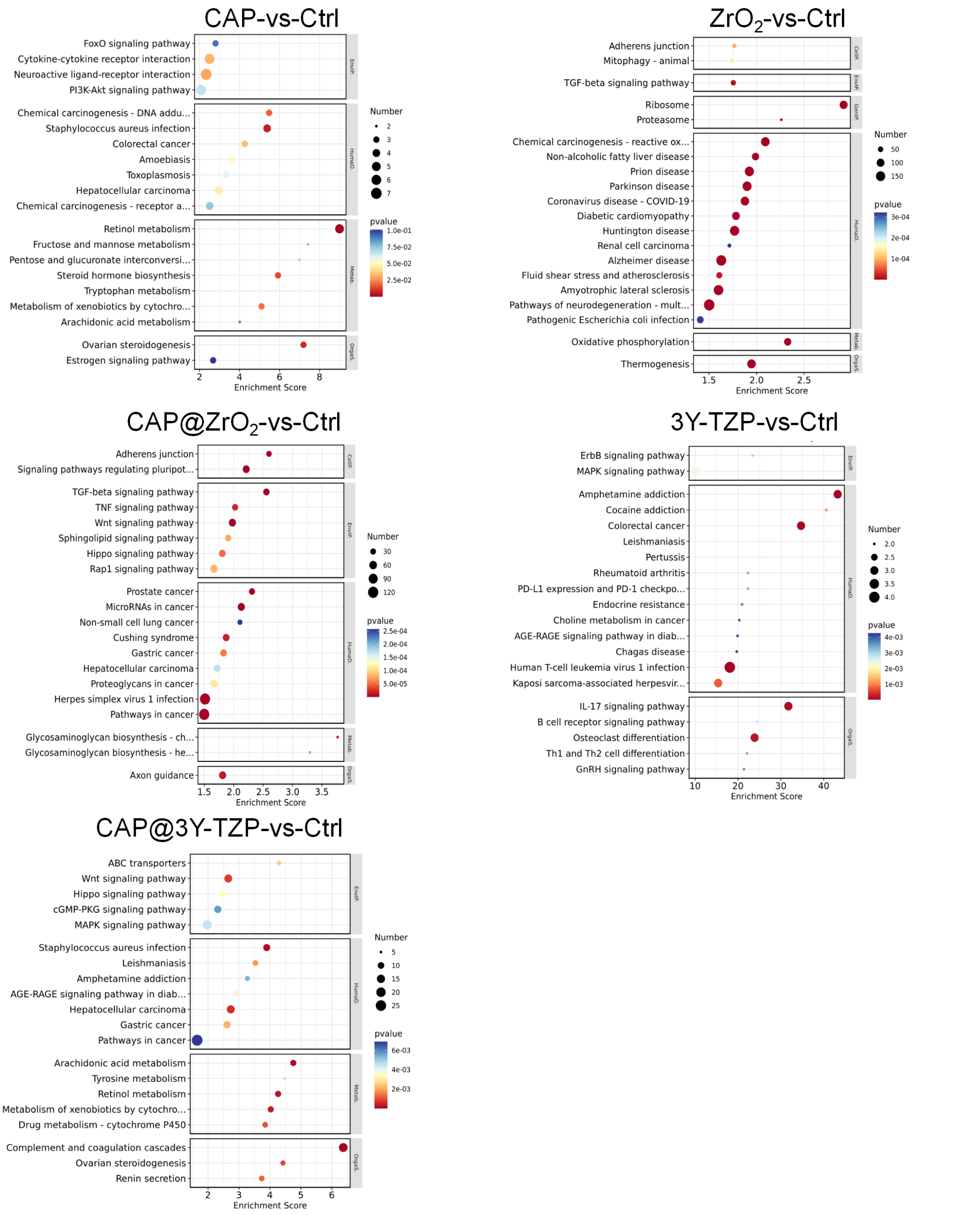

Fig. S9. Comparison of corresponding Reactome enriched pathways between different groups top20 bubble chart: CAP-vs-Ctrl, $ZrO_2$-vs-Ctrl, CAP@$ZrO_2$-vs-Ctrl, 3Y-TZP-vs-Ctrl, and CAP@3Y-TZP-vs-Ctrl.

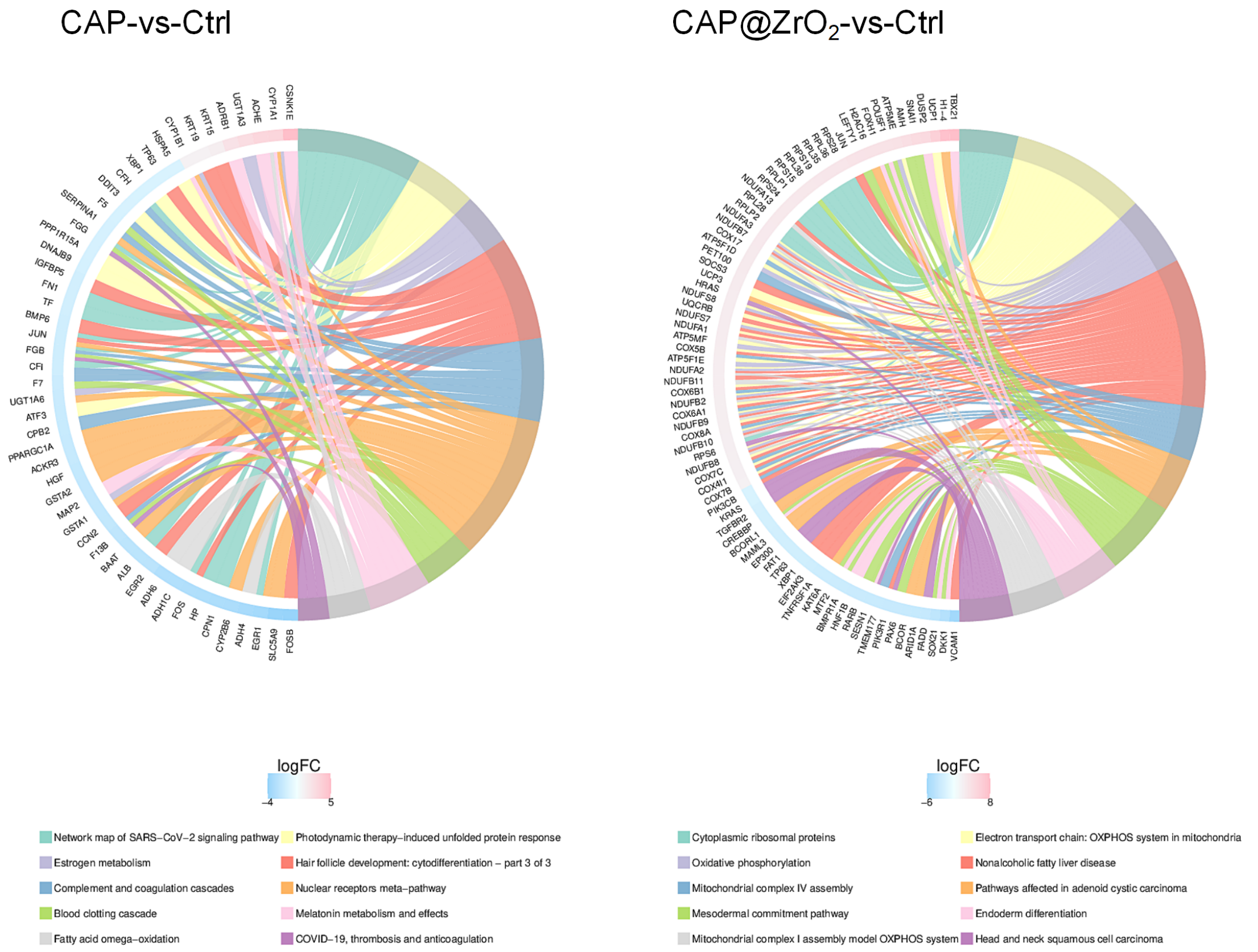

Fig. S10. WikiPathways enriched pathways corresponding to CAP-vs-Ctrl and CAP@$ZrO_2$-vs-Ctrl top10 chord diagram.

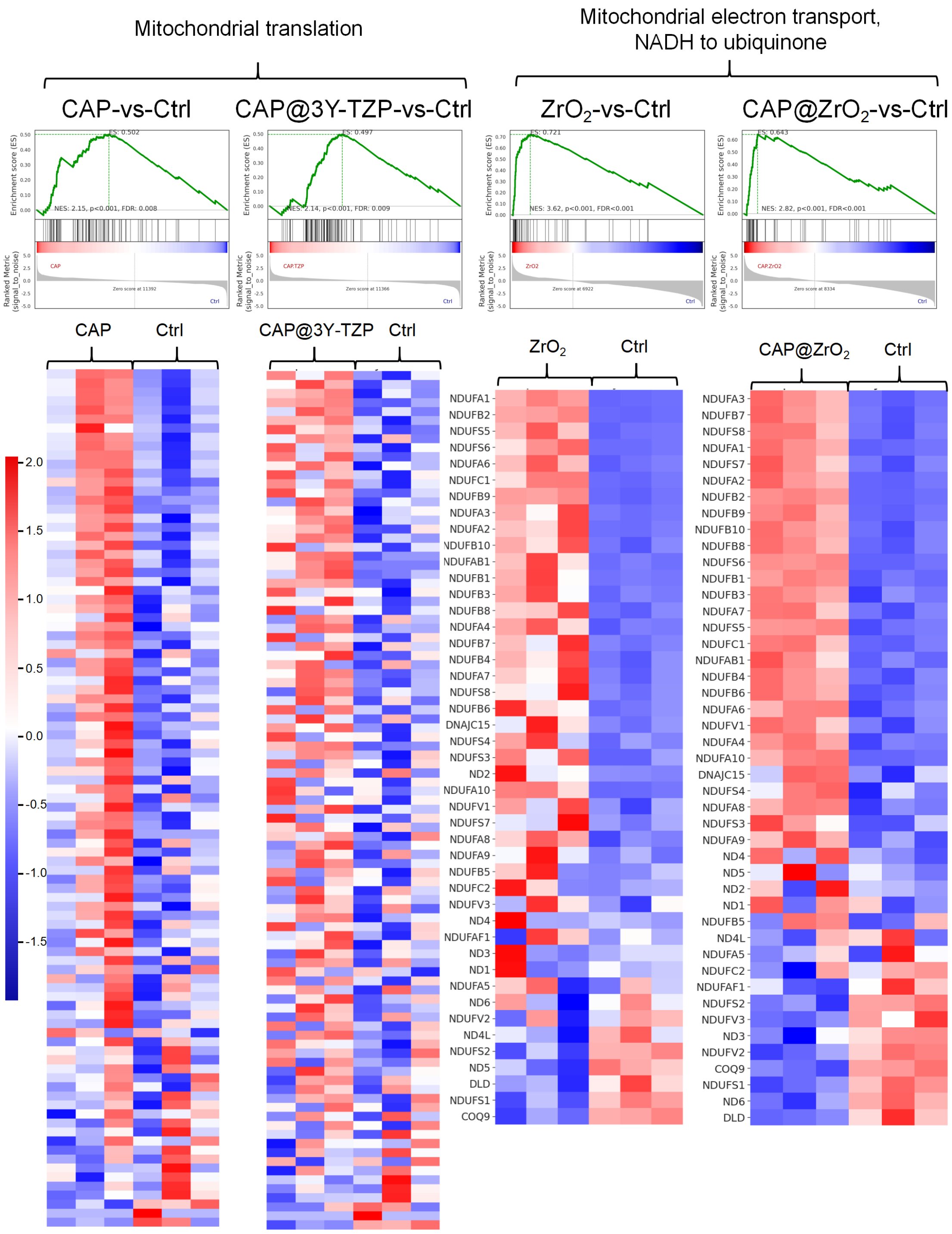

Fig. S11.The upper section displays Gene Set Enrichment Analysis (GSEA) enrichment analysis results for mitochondrial-related signaling pathways corresponding to the following comparisons: CAP-vs-Ctrl, CAP@3Y-TZP-vs-Ctrl, $ZrO_2$-vs-Ctrl, and CAP@$ZrO_2$-vs-Ctrl. Below, a heatmap visualizes the results of the GSEA analysis.

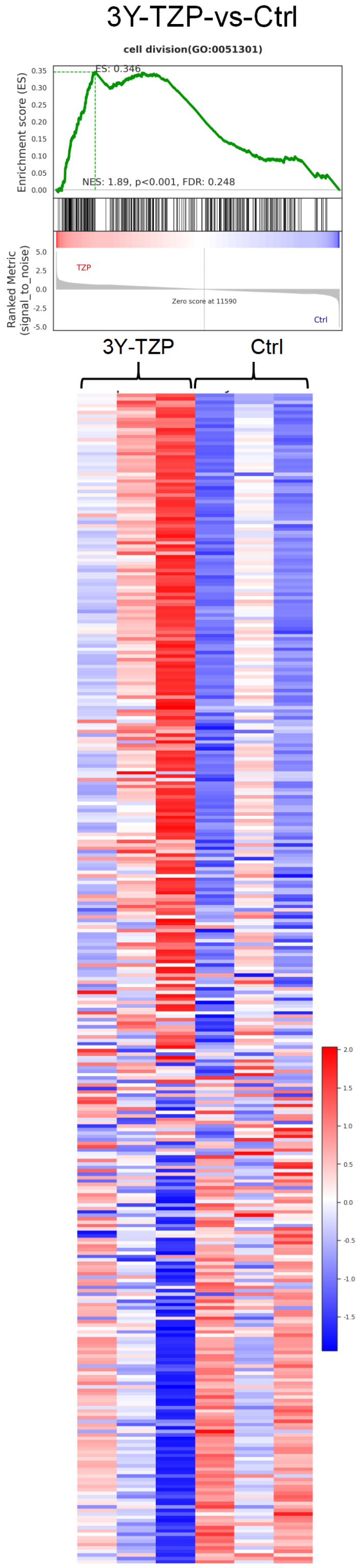

Fig. S12. The upper section of the figure illustrates Gene Set Enrichment Analysis (GSEA) enrichment analysis results for cell division-related signaling pathways in the comparison of 3Y-TZP-vs-Ctrl. Below, a heatmap visualizes the results of the GSEA analysis.

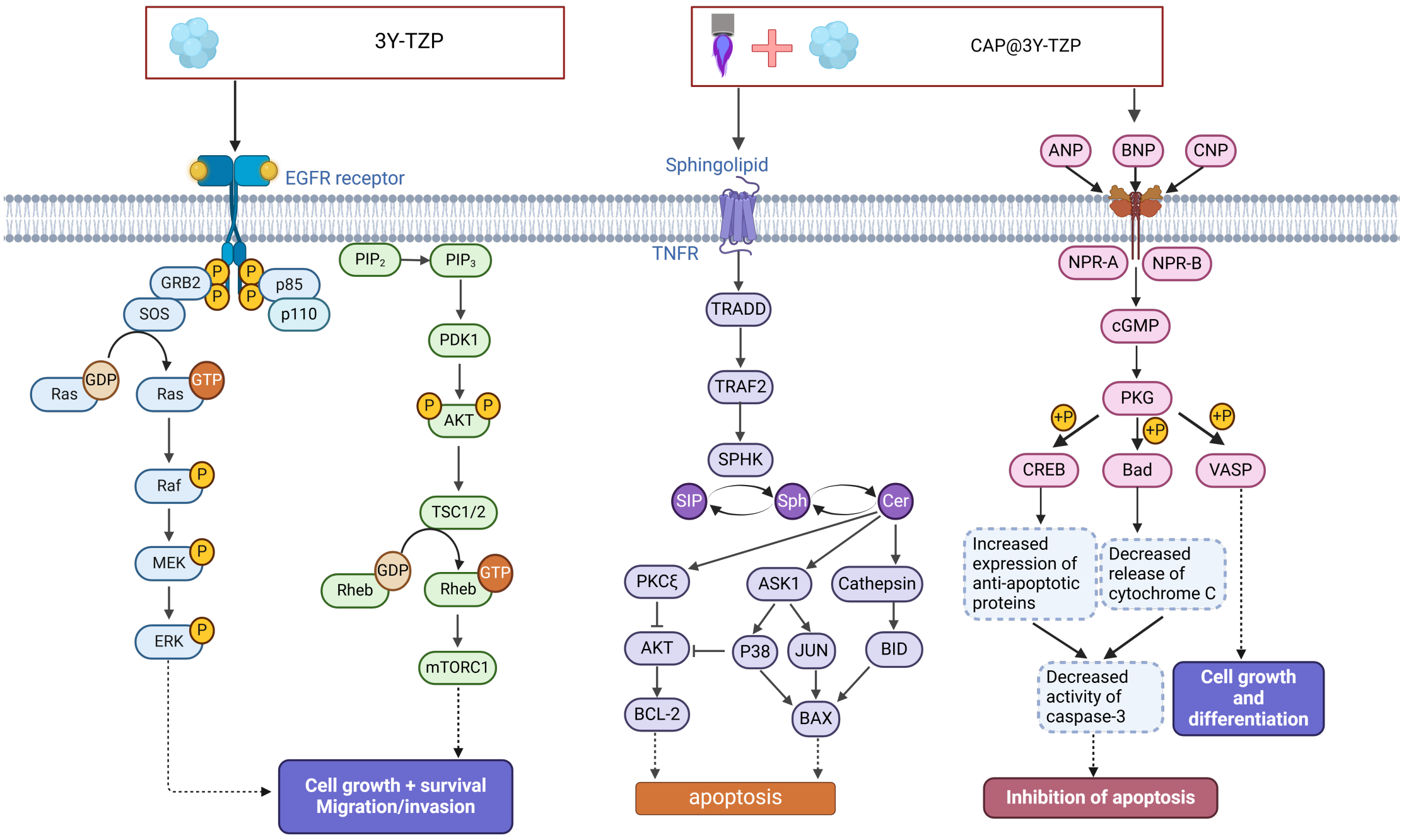

Fig. S17. The signaling pathway potential mechanism of 3Y-TZP NPs and CAP@3Y-TZP.

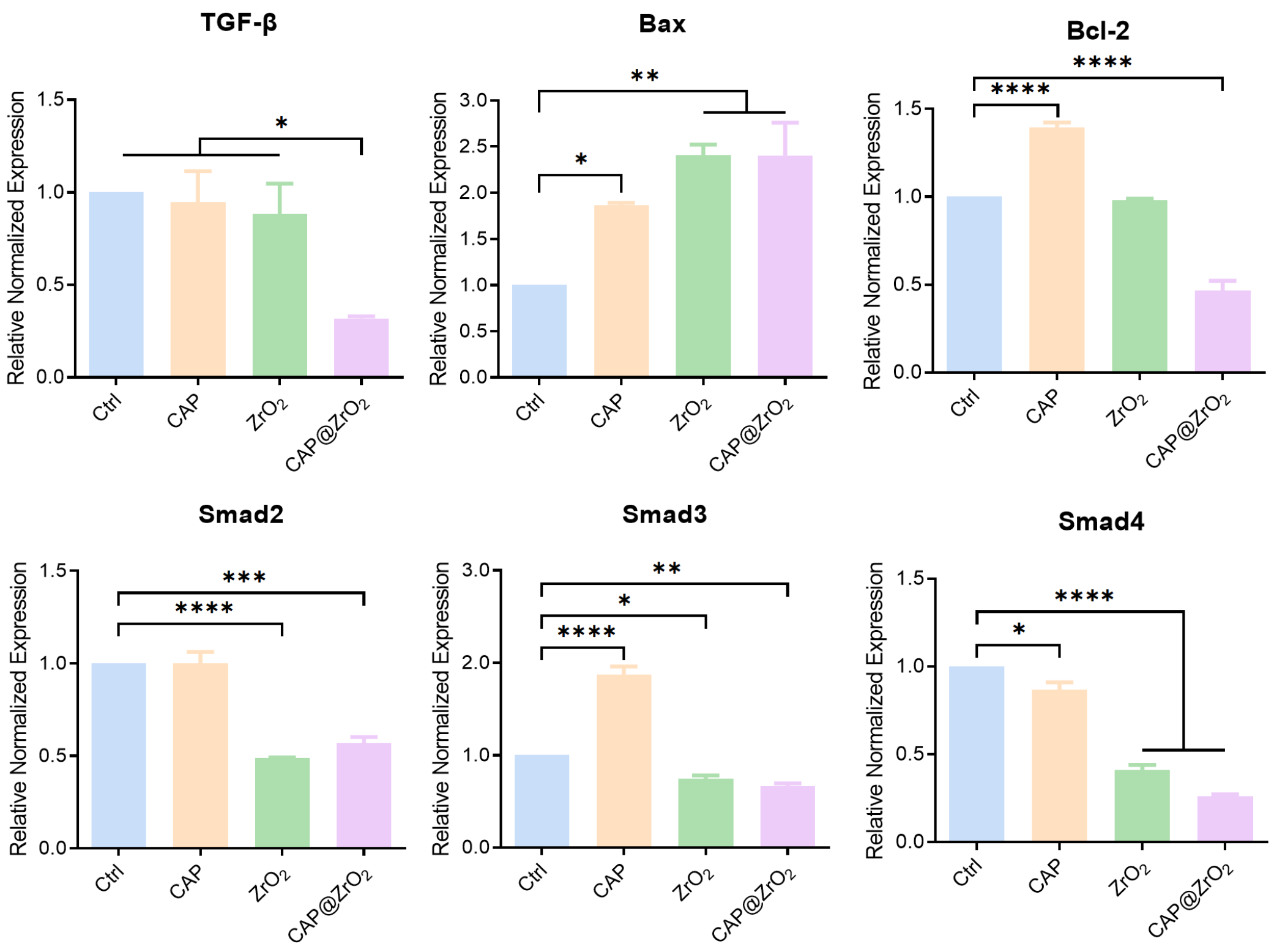

Fig. S18. TGF-β, Bax, Bcl-2, Smad2, Smad3, Smad4 transcript levels were measured by RT-qPCR.